\documentclass[12pt]{article}

\usepackage[english]{babel}
\usepackage{graphicx}
\usepackage{amsfonts}
\usepackage{amsmath}

\usepackage[title,titletoc]{appendix}
\usepackage[footnotesize,bf]{caption}
\usepackage{authblk}
\usepackage{setspace}
\usepackage[left=3cm,top=3cm,bottom=3cm,right=3cm,nohead,nofoot]{geometry}

\begin{document}

\title{Analogue neural networks on correlated random graphs}


\author[1,2]{Elena Agliari}
\author[3,4]{Lorenzo Asti}
\author[4,5]{Adriano Barra}
\author[1,2]{Raffaella Burioni}
\author[1,2]{Guido Uguzzoni}
\affil[1]{\footnotesize Dipartimento di Fisica, Universit\`{a} di Parma, viale G. Usberti 7/A, 43100 Parma, Italy.
\affil[2]INFN, Gruppo Collegato di Parma, viale G. Usberti 7/A, 43100 Parma, Italy.}
\affil[3]{Dipartimento di Scienze di Base e Applicate per l'Ingegneria, Sapienza Universit\`{a} di Roma, Via Antonio Scarpa 16, 00161 Roma, Italy.}
\affil[4]{Dipartimento di Fisica, Sapienza Universit\`{a} di Roma, Piazzale A. Moro 2, 00185 Roma, Italy.}
\affil[5]{GNFM, Sezione di Roma1}


\maketitle

\begin{abstract}
We consider a generalization of the Hopfield model, where the entries of patterns are Gaussian and diluted. We focus on the high-storage regime and we investigate analytically the topological properties of the emergent network, as well as the thermodynamic properties of the model. We find that, by properly tuning the dilution in the pattern entries, the network can recover different topological regimes characterized by peculiar scalings of the average coordination number with respect to the system size. The structure is also shown to exhibit a large degree of cliquishness, even when very sparse. Moreover, we obtain explicitly the replica symmetric free-energy and the self-consistency equations for the overlaps (order parameters of the theory), which turn out to be classical weighted sums of ``sub-overlaps'' defined on all possible sub-graphs. Finally, a study of criticality is performed  through a small-overlap expansion of the self-consistencies and through a whole fluctuation theory developed for their rescaled correlations: both approaches show that the net effect of  dilution in pattern entries is to rescale the critical noise level at which ergodicity breaks down.
\end{abstract}



\newcommand{\beq}{\begin{equation}}
\newcommand{\eeq}{\end{equation}}
\newcommand{\bea}{\begin{eqnarray}}
\newcommand{\eea}{\end{eqnarray}}
\newcommand{\np}{n_{\bf p}({\bf r},\,t)}
\newcommand{\vp}{{\bf v}_{\bf p}}
\newcommand{\ep}{\epsilon_{\bf p}}
\newcommand{\primo}{^\prime}
\newcommand{\sigsig}{({\bm \sigma}_1 \cdot {\bm \sigma}_2)}
\newcommand{\tautau}{({\bm \tau}_1 \cdot {\bm \tau}_2)}
\def\openone{\leavevmode\hbox{\small1\kern-3.8pt\normalsize1}}
\def\<{\langle}
\def\>{\rangle}
\def\lsim{\buildrel < \over {_{\sim}}}
\def\gsim{\buildrel > \over {_{\sim}}}
\def\dslash{\partial\!\!\!/}
\def\slash{\!\!\!/}
\def\sla{\!\!\!\!/}
\def\E{\mathbb{E}}
\def\ap{\left(\frac{1+a}{2}\right)}
\def\am{\left(\frac{1-a}{2}\right)}
\def\sumc{\sum_{\{\sigma\}} \int \prod_{\mu} d\mu(z_\mu)}
\def\sumimu{\sum_{i,\mu}^{l_{\eta},l_{\chi}}}
\def\sumi{\sum_{i}^{l_{\eta}}}
\def\summu{\sum_{\mu}^{l_{\chi}}}
\def \ap{\bigg(\frac{1+a}{2}\bigg)}
\def \g{\frac{\gamma}{2}}
\def \sg{\sqrt{\frac{\gamma}{2}}}

\section{Introduction}

In the last two decades, the application of statistical mechanics for the investigation of complex systems has attracted a growing interest, also due the number of applications ranging from economic \cite{coolen} and social sciences \cite{durlauf}, to biology and neurobiology \cite{amit}, to theoretical immunology \cite{JTB} and even to computer science \cite{marc} and machine learning \cite{ML}.
As a consequence, the introduction and the design of always deeper models along with the development of suitable techniques for their analysis becomes more and more
important for theoretical physicists and mathematicians involved in the field. In particular, attention has been devoted to the interplay of complex topology and critical behaviour, evidencing the strong relation between the network structures and the thermodynamics of statistical mechanics models \cite{doro}. 

In this paper we make a step forward in this direction, by analysing a Hopfield model \cite{hopfield,AGSa}, where pattern entries can be either extracted from a Gaussian distribution or set equal to zero. More precisely, entries are drawn from a normal distribution $\mathcal{N}[0,1]$ with a probability $(1+a)/2$ or set equal to zero with probability $(1-a)/2$, where $a \in [-1, +1]$ is a tunable parameter controlling the degree of dilution of patterns. We focus on the high-storage limit, namely the amount of patterns $L$ is linearly diverging with the system size $N$, i.e. $L = \alpha N$.

This kind of "analogue" neural networks has been intensively studied on fully connected topologies (see for instance \cite{bovier1, potts-hopf,a14, BGG1}) and further interest in the model lies in its peculiar "soft retrieval" as explained for instance in \cite{perezcastillo}.

Here, we first study the topological properties of the emergent weighted network, then we pass to the thermodynamic properties of the model.

In particular, we calculate analytically the average probability for two arbitrary nodes to be connected and we show that, by properly tuning $a$, the network spans several topological regimes, from fully connected down to the percolation threshold.
Moreover, even if the network is very sparse, it turns out to display a large degree of cliquishness due to the Hebbian rule underlying its couplings.
The coupling distribution is also explicitly calculated and shown to be central and with extensive variance, as expected.

From a thermodynamic perspective, using an exact Gaussian mapping, we prove that this model is equivalent to a bipartite diluted spin-glass (or to a Restricted Boltzman Machine \cite{benjo} retaining a cognitive system perspective \cite{hotel}), whose parties are made up by binary Ising spins and by Gaussian spins, respectively, while interactions among them, if present, are drawn from a standard Gaussian distribution $\mathcal{N}[0,1]$; of course, there are no links within each party.
The size of the two parties are respectively $N$, for the Ising spins, and $L$, for the Gaussian ones, and the dilution in the Hopfield pattern entries corresponds to standard link removal in this bipartite counterpart.
\newline
Then, extending the technique of multiple stochastic stability (developed for fully connected Hopfield model in \cite{BGG1} and  for ferromagnetic systems on small-world graphs in \cite{SMnostro}) to this case, we solve its thermodynamics at the replica symmetric level. Once introduced suitably order parameters for this theory, we obtain an explicit expression for the free-energy density that we extremize with respect to them to obtain the self-consistencies that constraint the phase space of the model.
As in other works on diluted networks \cite{SMnostro, ton1, ton2}, the order parameters are two (one for each party) series  of overlaps defined on all the possible subgraphs through which the network can be decomposed.
A study of their rescaled and centered fluctuations allows to obtain the critical surface delimiting the ergodic phase from the spin-glass one. The same result is also recovered through small overlap expansion from the self-consistencies; the agreement confirms the existence of a second order phase transition \cite{amit,MPV}.

The paper is organized as follows: in Section $2$ the model is defined with all its related parameters and variables, while in Section $3$ its topological properties are discussed. Section $4$ deals with the statistical mechanics analysis while in Section $5$ fluctuation theory is developed. Section $6$ is left for a discussion and outlooks.

\section{The model}

Given $N$ Ising spins $\sigma_i = \pm 1$, $i \in (1, ..., N)$, we aim to study a mean-field model whose Hamiltonian has the form
\beq\label{eq:Hamiltonian_D}
\tilde{H}= - \frac{1}{D} \sum_{ij}^{N} J_{ij} \sigma_i \sigma_j
\:,
\eeq
where the couplings are built in a Hebbian fashion \cite{hebb}\cite{hopfield} as
\beq\label{eq:J_tilde}
J_{ij} = \sum_{\mu=1}^L \xi_i^\mu \xi_j^\mu,
\eeq
and $D$ is a denominator whose specific form is discussed in Section $3$. In fact, in general, as the coordination number may vary sensibly according to the definition of patterns $\xi$, in order  to ensure a proper linear scaling of the Hamiltonian (\ref{eq:Hamiltonian_D}) with the volume, $D$ has to be a function of the system size $N$ and of the parameters through which patterns $\xi$ are defined.

We consider the high-storage regime \cite{peter}, such that, in the thermodynamic limit (i.e. $N \to \infty$), the following scaling for the amount of stored memories (patterns) is assumed
\beq
\lim_{N\rightarrow\infty} \frac{L}{N} = \alpha \in \mathbb{R}^+,
\eeq
\
 even though we use the symbol $\alpha$ for the ratio between the number of patterns and the system size also at finite $N$, bearing in mind that the thermodynamic limit has to be performed eventually.

The quenched entries of the memories $\xi_i^{\mu}$ are Gaussian and diluted, namely they are set to zero with probability $(1-a)/2$, while, with probability $(1+a)/2$, they are drawn from a standard Gaussian distribution:
\beq
 P(\xi_i^{\mu})=
 \am \delta(\xi_i^\mu) \,+\, \ap \mathcal{N}_{[0,1]}(\xi_i^\mu)
 \:.
\eeq
The parameter $a$ can in principle be varied in the range $a \in [-1,1]$, and, in general, small values correspond to highly diluted regimes. As proved in Section \ref{subsec:topology}, a scaling law for this parameter has to be introduced in order to avoid the topology to become trivial in the thermodynamic  limit. Thus, we consider the following scaling
\beq\label{eq:scaling}
a = -1 +\frac{\gamma}{N^{\theta}}
\:,
\eeq
where $\theta$ determines the topological regime of the network, while $\gamma$ plays the role of a fine tuning within it. More precisely, $\gamma \in (0,2]$ and, of course, for $\gamma=0$ we get $ P(\xi_i^{\mu})= \delta(\xi_i^\mu)$, that is, there is no network, so we  discuss only the case $\gamma>0$. Finally, notice that fixing $\theta=0$ and $\gamma=2$ yields to $a=1$, corresponding to the standard analogue Hopfield model \cite{BGG1}.

\section{Topological analysis}\label{sec:topology}

\subsection{Coupling distribution}
\begin{figure}[htbp]
\centering
\includegraphics[scale=.65]{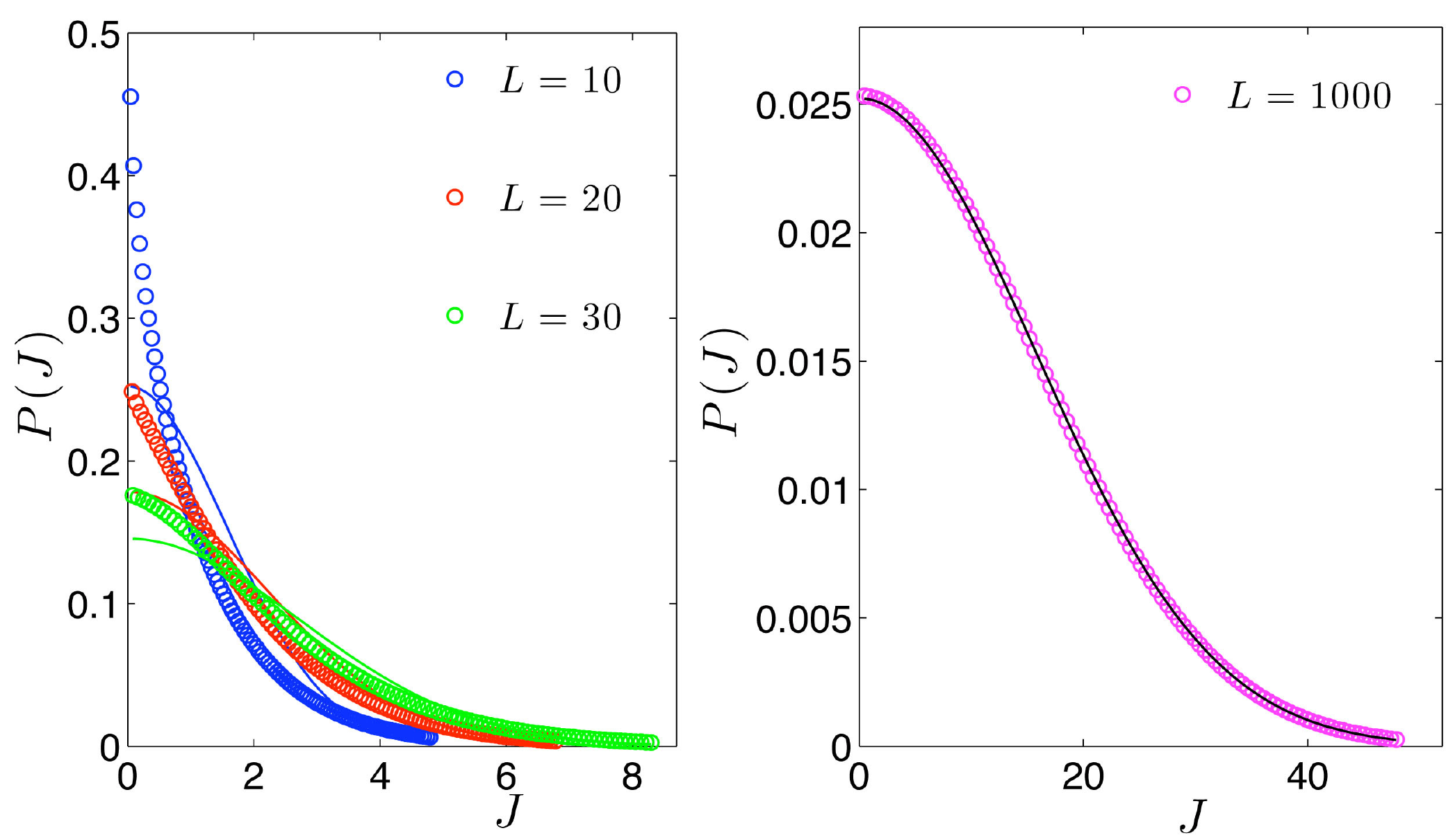}
\caption{Coupling distribution $P(J)$ for $L=10$ (left panel, blue), $L=20$ (left panel, red), $L=30$ (left panel, green) and $L=1000$ (right panel). Circles represent the coupling distribution calculated according to Eq.~\ref{eq:PJ1}, while the continuous lines represent normal distributions with momenta given by Eq.~\ref{eq:meanJ} and Eq.~\ref{eq:varJ}: as the thermodynamic limit is approached, the agreement gets better and better.
Notice that, when $L$ grows, the divergence in $J=0$ becomes weaker.
Only the positive values of $J$ are considered due to the symmetry.}
\label{fig:PJ}
\end{figure}

Let us consider the definition of the coupling strength in Equation~(\ref{eq:J_tilde}): the probability $p$ that the $\mu$-th term $\xi_i^\mu\xi_j^\mu$ is zero corresponds to the probability that at least one between $\xi_i^\mu$ and $\xi_j^\mu$ is zero, which is
\beq
p \equiv
\am^2 + 2 \am\ap= \frac{3-a^2-2a}{4}=1-\ap^2
\:,
\eeq
while its complement is the probability that a Gaussian number is drawn for both entries, that is $(1-p) = [(1+a)/2]^2$.
Thus, the probability that the link connecting $i$ and $j$ has strength $J_{ij}$ can be written as
\bea
P(J_{ij}) &&=
p^L  \delta(J_{ij}) +
\sum_{k=1}^L p^{L-k} (1-p)^{k} {L \choose k}~ P_k\big(\sum_{\nu}^k \xi_i^\nu \xi_j^\nu = J_{ij}\big) =\label{eq:PJ1}\\
 &&= p^L  \delta(J_{ij}) + \sum_{k=1}^L f(k) ~ P_k(\sum_{\nu}^k \xi_i^\nu \xi_j^\nu = J_{ij}) =\nonumber \\
 && = p^L  \delta(J_{ij}) + \sum_{k=1}^L f(k) ~ \int_{-\infty}^{+\infty}\frac{dl}{2\pi} \frac{e^{-il J_{ij}}}{\left( 1 + l^2\right)^{k/2}}\label{eq:PJ2}
\:,
\eea
where to simplify the notation we defined $f(k)=p^{L-k} (1-p)^{k} {L \choose k}$ and $P_k\big(\sum_{\nu}^k \xi_i^\nu \xi_j^\nu = J_{ij}\big)$ is the probability that $k$ pairs of Gaussian entries, pairwise multiplied, sum up to $J_{ij}$, namely
\bea
&&\!\!\!\!\!\!\!\!\!P_k\bigg(\sum_{\nu}^k \xi_i^\nu \xi_j^\nu = J_{ij}\bigg) =
\int_{-\infty}^{\infty} \prod_{\nu=0}^k d\xi_i^\nu d\xi_j^\nu
\, P(\xi_i^\nu) P(\xi_j^\nu) \, \delta\bigg(\sum_{\nu=0}^k \xi_i^\nu \xi_j^\nu - J_{ij}\bigg) =
\nonumber\\
&&\!\!\!\!\!\!\!\!\!=
\int_{-\infty}^{+\infty} \frac{dl}{2\pi} \prod_{\nu=0}^k d\xi_i^\nu d\xi_j^\nu
\, \frac{e^{-\frac{(\xi_i^\nu)^2}{2}}}{\sqrt{2\pi}} \frac{e^{-\frac{(\xi_i^\nu)^2}{2}}}{\sqrt{2\pi}}
\,e^{il\left(\xi_i^\nu \xi_j^\nu - J_{ij}\right)} =
\int_{-\infty}^{+\infty} \frac{dl}{2\pi} \frac{e^{-ilJ_{ij}}}{\left( 1 + l^2\right)^{k/2}}
\:.
\eea
From Equation (\ref{eq:PJ2}) one can easily specify the characteristic function of the coupling distribution
\beq
F(l) \equiv \int_{-\infty}^{+\infty} e^{i l J} P(J) dJ = p^L+ \sum_{k=1}^L  \frac{ f(k)}{\left( 1 + l^2\right)^{k/2}}=\frac{1}{2} p^L\left[1+\left(\frac{1+l^2 p}{p+l^2 p}\right)^L\right]
\:,
\eeq
where we dropped the indices $i$ and $j$, due to the arbitrariness of the couple of nodes considered.
From $F(l)$ it is possible to obtain all the momenta by simple differentiation. For instance, first and second moment read respectively as
\beq\label{eq:meanJ}
\E[J_{ij}]= (-i) \frac{\partial F(l)}{\partial l} \bigg|_{l=0}= 0,
\eeq
\beq\label{eq:varJ}
\E[J_{ij}^2]=(-i)^2 \frac{\partial^2 F(l)}{\partial l^2} \bigg|_{l=0}= L (1-p)= L\ap^2=\frac{\alpha \gamma^2}{4} N^{1-2\theta}
\:.
\eeq
Now, for fixed $a$ and $\alpha$, we expect that $J$, being a sum of Gaussian variables, is also normally distributed (except the point $J=0$), at least for large $N$. Indeed, numerical simulations confirm that the distribution $P(J)$ converges in the thermodynamic limit ($L\rightarrow \infty$) to a Gaussian distribution with  zero mean and variance given by Equation (\ref{eq:varJ}) (see Figure \ref{fig:PJ}), except for the point $J=0$ which will be discussed in the following section.

\subsection{Link Probability and topology regimes}\label{subsec:topology}
Let us consider the bare topology. The quantity of interest is the average link probability $P_{\textrm{link}}$:
\beq\label{eq:Plink}
P_{\mathrm{link}}= 1 - P(J=0)
\:.
\eeq

Looking at Equation (\ref{eq:PJ1}), in principle $P(J=0)$ has two contributions: one from the delta function and one from the sum over $k$ random numbers, but the latter has a null measure in the limit $L \rightarrow \infty$. To show this we consider the second term in Equation (\ref{eq:PJ1}) and we calculate its measure over the interval $J_{ij}\in[-\epsilon,+\epsilon]$, highlighting for clarity the term $k=1$:
\bea
&&\!\!\!\!\!\!\!\!\!\!\int_{-\epsilon}^\epsilon dJ_{ij} \sum_{k=1}^L f(k) ~ P_k(\sum_{\nu}^k \xi_i^\nu \xi_j^\nu - J_{ij})= \nonumber\\
&&\!\!\!\!\!\!\!\!\!\!=p^{L-1} ~(1-p) L \int_{-\epsilon}^\epsilon P_1(r) dr~+~\sum_{k=2}^L  ~{L \choose k}~ p^{L-k} ~(1-p)^{k}~  \int_{-\epsilon}^\epsilon P_k(r) d r \label{eq:14}
\:.
\eea
In fact, we notice that $P_1(r)$ has a weak divergence in $r=0$ and its integral scale as $\sim \epsilon  \log(\epsilon)$, so that the divergence is suppressed by the prefactor in the limit $L \rightarrow \infty$, so that the first term in Eq.~\ref{eq:14} is vanishing. As for $P_{k>1}(r)$, its integral is non-diverging and can be upper bounded \footnote{
From the the two inequalities $ \int_{-\epsilon}^\epsilon P_k(r) dr < 2\epsilon P_k(0), \quad P_k(0)<\mathrm{c} \frac{\log(k)}{k}$ (with $c$ a constant), it follows
$$\sum_{k=2}^L  ~ {L \choose k}~ p^{L-k} (1-p)^k~  \int_{-\epsilon}^\epsilon P_k(r) dr < 2  \epsilon \mathrm{c}, \sum_{k=2}^L {L \choose k}~ p^{L-k} ~(1-p)^k \frac{\log(k)}{k}$$
which goes to zero in the limit $L \rightarrow \infty$.
} to show that the second term is also negligible in the limit $L \rightarrow \infty$.

Hence, in the thermodynamic limit, $J_{ij}=0$ only if $\xi_i^{\mu} \xi_j^{\mu}=0$, for any $\mu$, namely
\beq\label{e:P(J=0)}
P(J_{ij}=0) =p^L=\left( \frac{3-a^2-2a}{4} \right)^L
 \:.
\eeq

Now, looking at (\ref{eq:Plink}) and (\ref{e:P(J=0)}) in the thermodynamic limit, it is clear that, if we consider $a$ as finite and constant, only two trivial topologies can be realized. In fact, if  $a= -1$, $P_{\mathrm{link}}$ is zero and the system is fully disconnected, while, with $a > -1$, $P_{\mathrm{link}}$ tends to one exponentially fast with the system size, and the graph becomes fully connected.

Nevertheless, with the scaling (\ref{eq:scaling}),
\beq\label{eq:Plink2}
P_{\textrm{link}} = 1- \left( 1-\frac{\gamma^2}{4N^{2\theta}} \right)^{\alpha N}
\simeq 1 - e^{-\frac{\alpha\gamma^2}{4N^{2\theta-1}}},
\eeq
where the last expression holds for large $N$ and $\gamma \in (0,2]$. Now, by tuning the value of $\theta$, we realize different topological regimes; within each regime the parameter $\gamma$ acts as a fine tuning. Following a mean-field approach, namely just focusing on the average link probability, we can distinguish:

\begin{itemize}
	\item \underline{$\theta =0$} : $P_{\mathrm{link}} = 1- \left( 1-\frac{\gamma^2}{4}\right)^{\alpha N} \rightarrow 1$.
	Fully Connected graph, with average degree equal to the system size ($\bar{z}=N-1$).\\
	The coupling distribution converges to the Gaussian one with variance $\mbox{Var}[J] \propto N$, as in the with Sherrington-Kirkpatrick model.
	\item \underline{$0 < \theta < 1/2$}: $P_{\mathrm{link}} = 1 - e ^{-\frac{\alpha\gamma^2}{4}N^k} \rightarrow 1$ (where $0<k<1$).
	Fully Connected graph, with average degree equal to the system size ($\bar{z}=N-1$).\\
	The coupling distribution converges to the Gaussian one with $\mbox{Var}[J] \propto N^{k}$.\\
	\item \underline{$ \theta = 1/2$}: $P_{\mathrm{link}} \simeq \frac{\alpha \gamma^2}{4} = \textrm{const}.$ The link probability is finite and the average coordination number is linearly diverging with the system size, namely $\bar{z} = \alpha \gamma^2 (N-1) = \mathcal{O}(N)$, and $\mbox{Var}[J] \propto \mbox{cost}$.
	\item \underline{$1/2 < \theta < 1$}: 	$P_{\mathrm{link}} = 1 - e ^{-\frac{\alpha\gamma^2}{4}N^{k}} \simeq \frac{\alpha\gamma^2 N^k}{4} \rightarrow 0$ (where $-1<k<0$). Extreme Diluted Graph, characterized by a sublinearly diverging average coordination number, $\bar{z} = \mathcal{O}(N^{1-k})$, and $\mbox{Var}[J] \propto N^{k}$.\
	\item \underline{$\theta = 1$}: $P_{\mathrm{link}} = 1 - e ^{-\frac{\alpha\gamma^2}{4N}} \simeq \frac{\alpha\gamma^2}{4N}\rightarrow 0$. Finite Coordination Regime with $\bar{z}=  \alpha \gamma^2/4$, and $\mbox{Var}[J] \propto 1/N$.
	\item \underline{$\theta > 1$}: $P_{\mathrm{link}} = 1 - e ^{-\frac{\alpha\gamma^2}{4}N^{k}} \simeq \frac{\alpha\gamma^2 N^k}{4} \rightarrow 0$	(where $k<-1$). Fully Disconnected Regime with coordination number vanishing for any choice of $\alpha$ and $\gamma$. The variance of the coupling distribution is vanishing superlinearly with $N$.
	\end{itemize}

\begin{figure}[htbp]
\centering
\includegraphics[scale=.6]{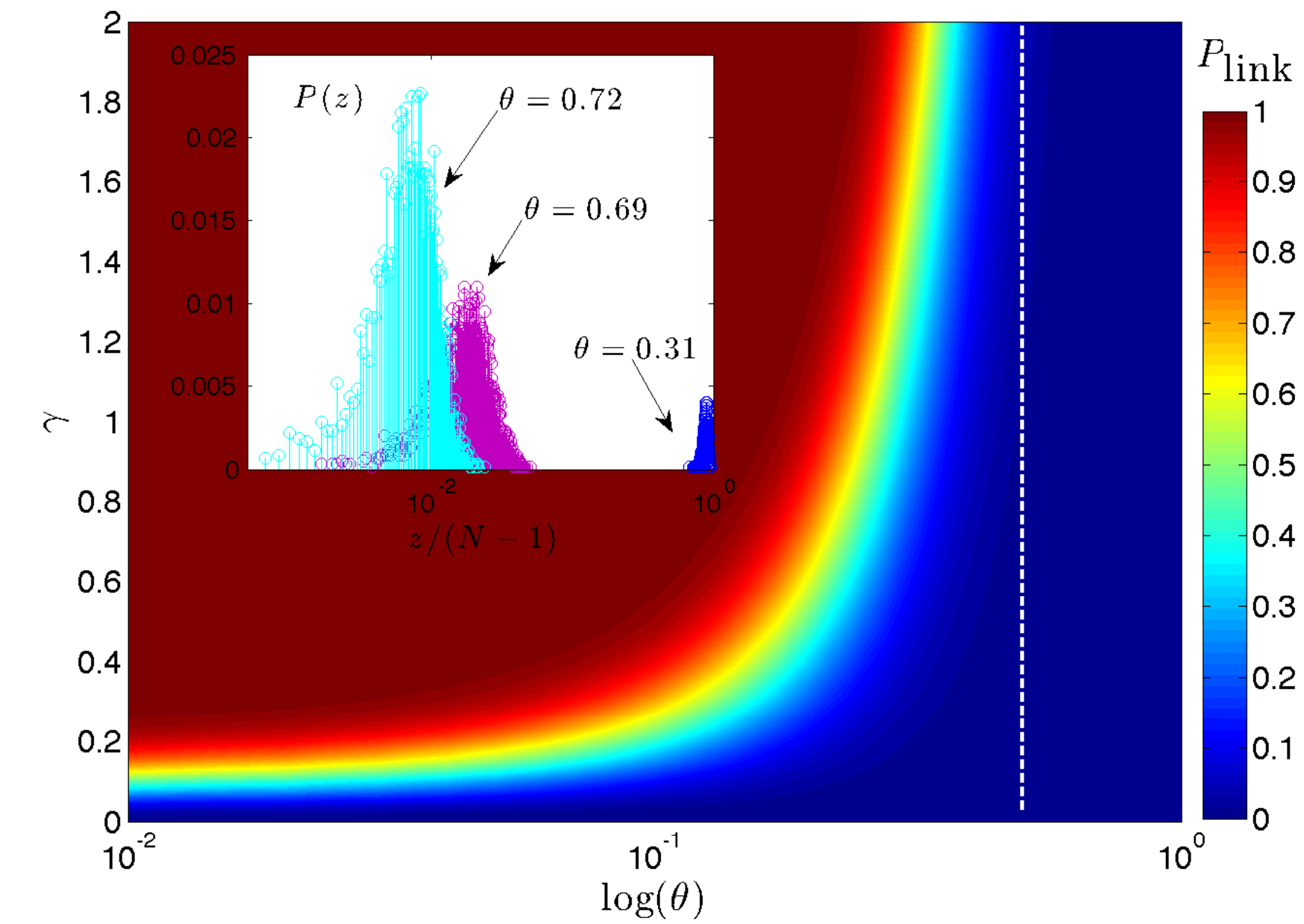}
\caption{The main figure represents the contour plot of $P_{\textrm{link}}$ (see Equation~\ref{eq:Plink2}) as a function of $\gamma$ and of $\log \theta$. The dashed, vertical line corresponds to $\theta=1/2$ and demarcates the onset of a disconnected regime. The inset represents the degree distribution $P(z)$ as a function of the normalized number of nearest neighbors; three values of $\theta$ are considered as specified. Notice that, as expected, larger values of $\theta$ yields to sparser graphs. Both figures refer to systems made up of $N=6000$ nodes, with $\alpha=0.05$ and $\gamma=1$.}
\label{fig:Plink}
\end{figure}

A contour plot of $P_{\textrm{link}}$ as a function of $\gamma$ and $\theta$ is shown in Figure \ref{fig:Plink}.

\subsection{Small-world properties}
Small-world networks are characterized by two main properties:  a small diameter and a large clustering coefficient, namely, the average shortest path length scales logarithmical (or even slower) with the system size and they contain more cliques than what expected by random chance \cite{strogatz}.
The small-world property has been observed in a variety of real networks, including biological and technological ones \cite{mendes}.

First, we checked that, in the overpercolated regime, the structures
considered here display a diameter growing logarithmically with $N$,
as typical for random networks \cite{doro}. 

As for the clustering coefficient $C$, it is basically defined as the likelihood that two neighbors of a node are linked themselves, that is, for the $i$-th node,
\beq
c_i=\frac{2 E_i}{z_i(z_i-1)}
\:,
\eeq
where $z_i$ is the number of nearest neighbors of $i$ and $E_i$ is the number of links connecting any couple of neighbors; when $E_i$ equals its upper bound $z_i(z_i-1)/2$, the neighborhood of $i$ is fully connected.
The global clustering coefficient then reads as
\beq
C= \frac{1}{N} \sum_i^N c_i.
\eeq
A clustering coefficient close to 1 means that the graph displays a high ``cliquishness", while a value close to 0 means that there are few triangles.

\begin{figure}[htbp]
\centering
\includegraphics[scale=.7]{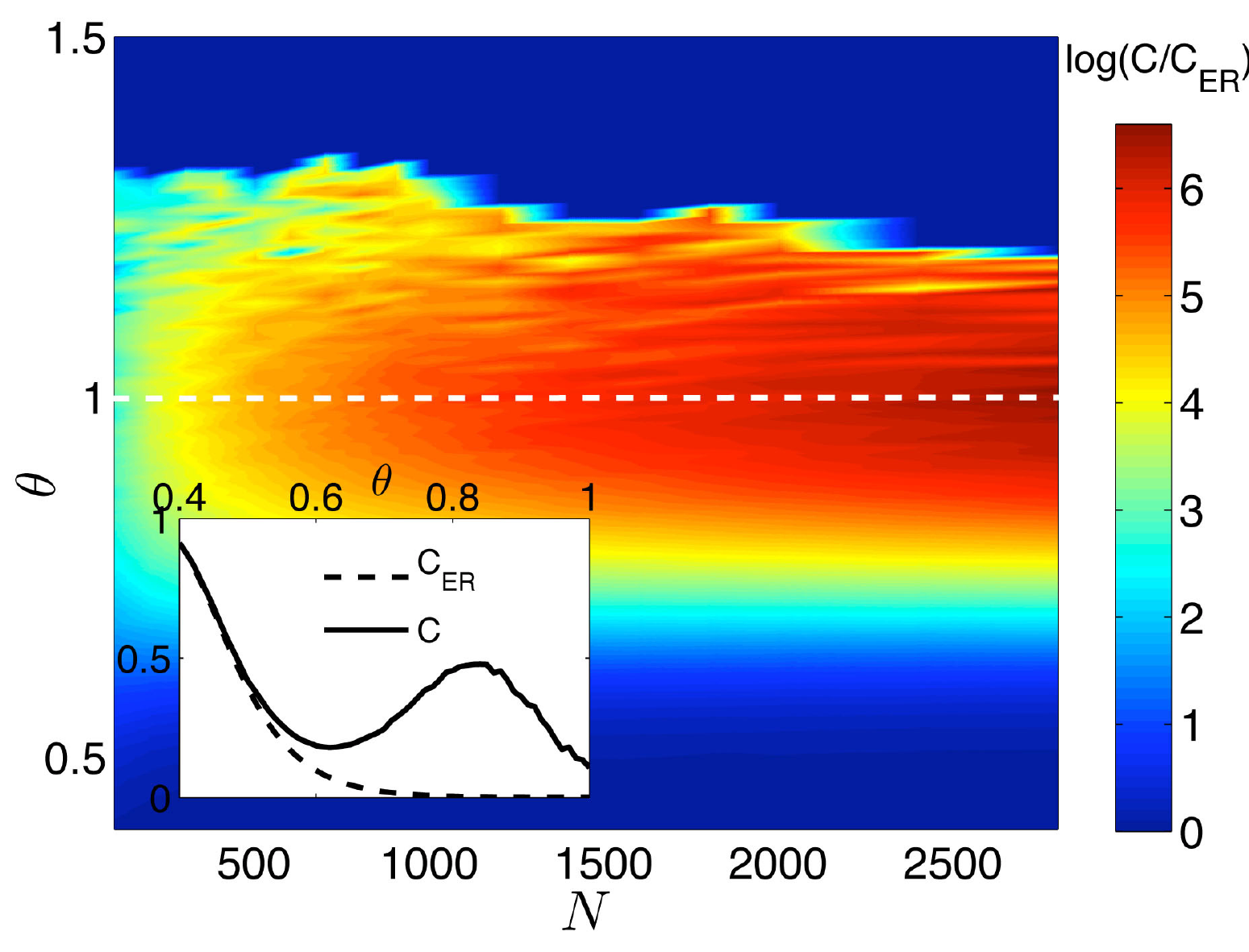}
\caption{Inset: Mean clustering coefficient $C$ (continuous line) for different choices of the parameter $\theta$, while $N=1600$,
$\gamma=2$ and $\alpha=0.5$ are kept fixed; the mean has been performed over all nodes making up the graph and over $10^2$ realizations. An analogous ER random graph is also considered and the related clustering coefficient $C_{\textrm{ER}}$ (dashed line) is shown for comparison. Notice the qualitative different behaviors of $C$ and $C_{\textrm{ER}}$.
Main figure: contour plot for the logarithm of the ratio $C/C_{\textrm{ER}}$, as a function of $N$ and $\theta$. Notice that, although for $\theta$ close to $1$ both graphs are sparse, $C \gg C_{\textrm{ER}}$. On the other hand, for $\theta >1$, both coefficient converge to zero, in the thermodynamic limit; the non-null values appearing in the figure are due to finite-size effects.
}
\label{fig:c_mean}
\end{figure}

It is easy to see that for the Erd\H{o}s-R\'enyi graph, where each link is independently drawn with a probability $P$, the average clustering coefficient is
$C_{\textrm{ER}} = P$. Therefore, for our network, we measure $C$ and we compare it with the average link probability $P_{\textrm{link}}$; results obtained for different choices of $\theta$ are shown in Figure \ref{fig:c_mean}.

First, we notice that, for a given system size $N$, the behavior of $C$ and of $C_{\textrm{ER}}$, with respect to $\theta$, is markedly different (see the inset): the latter decreases monotonically due to the analogous decrease of the link probability, while the former exhibits two extremal points at a relatively large degrees of dilution. In fact, as long as the networks are highly connected, the disappearance of a few links yields, in both cases, a modest drop in the overall cliquishness. On the other hand, when dilution is significant, the intrinsic structure of the ``Hebbian graph'' matters: as patterns get sparser and sparser, surviving links are those connecting nodes whose related patterns display matching with non-null entries. In this way, the neighbors of a node are also likely to be connected \cite{agliari2,SMnostro,cioli,cioli2} and the clustering coefficient grows. Finally, at a very large degree of dilution, the system approaches the fully-disconnected regime and the clustering coefficient decreases.

In order to compare more effectively our graph and an analogous ER graph, we also considered the ratio $C/C_{\mathrm{ER}}$ (see the main figure). Interestingly, for $\theta$ relatively large, as $N$ gets larger this ratio grows confirming that the few links remaining are very effective in maintaining the cliques.
This can be understood as follows: to fix ideas let us take $\theta=1$, so that the average number of non-null entries in a string is $L \gamma/ (2N)$ which equals $\gamma \alpha /2$ in the high-storage regime under investigation. For simplicity, let also assume that  $\gamma \alpha /2 \approx 1$, and that this holds with vanishing variance for all nodes. Therefore, if the node $i$ has $k$ neighbors, its (local) clustering coefficient is either $0$ (if $k \leq 2$) or $1$ (if $k>2$).  Hence, the expected local clustering coefficient can be estimated as the probability for a node to display $k>2$ nearest neighbors, namely $1 - (1- q)^{N-1} - Nq(1-q)^{N-2}$, where $q = \gamma / (2N)$ is the probability that the pattern of an arbitrary node $j \neq i$ has the non-null entry matching with the one of $\xi_i$. With some algebra we get $c_i \approx 1 - e^{- \gamma/2} (1 + \gamma/2)$, which remains finite also in the thermodynamic limit, in agreement with results from simulations. For $\theta >1$, $q = \gamma N^{1- \theta}/ 2$ and $c_i$ converges to zero.

\section{The statistical mechanics analysis}\label{sec:mecstat}

In this section we study the thermodynamic properties of the system introduced: At first we show its equivalence to a bipartite spin-glass and figure out the order parameters of the theory, then we define an interpolating free-energy which generalizes the multiple stochastic stability developed in \cite{BGG1}; this technique allows to obtain the replica-symmetric solution in form of a simple sum rule. As a last step, we extremize the free energy finding self-consistencies for the order parameters, whose critical behavior is also addressed.

\subsection{The equivalent diluted bipartite spin-glass}
As we deal with a structure whose average coordination number may range in $[0, N]$, from a statistical mechanics perspective,  we aim to define the normalization constant $D$ for the Hamiltonian in Equation~(\ref{eq:Hamiltonian_D}), in such a way that its average (which defines the extensive energy of the system and is denoted symbolically with the brackets) is linearly diverging with the system size, namely $\< \tilde{H} \> \propto N$.

By  a direct calculation, it is possible to show that this condition is fulfilled by
\beq\label{eq:normalization}
D= N^{1-\theta},
\eeq
so that, using the explicit definition for the couplings, we can write
\beq\label{eq:Hamiltonian_}
\tilde{H}=
- \frac{1}{N^{1-\theta}} \sum_{i<j}^{N} \sum_{\mu}^L \xi_i^\mu \xi_j^\mu \sigma_i \sigma_j=
- \frac{1}{N^{1-\theta}} \sum_{i<j}^{N} J_{ij} \sigma_i \sigma_j
\:,
\eeq

For a single realization of the disorder encoded in the memories, the partition function reads off as:
\beq\label{eq:bipartite}
\tilde{Z}_{N,L}(\beta;\xi) =
\sum_{\left\{\sigma\right\}}
\;\mbox{exp} \left\{
\frac{\beta}{2N^{1-\theta}}
\sum_{i,j}^{N} \sum_{\mu}^{L} \xi_i^{\mu} \xi_j^{\mu} \sigma_i \sigma_j
\right\}
\:.
\eeq
Note that, as usual in the Hopfield model, the diagonal term gives an extensive contribution to the partition function. In the above expression we neglected this diagonal term, directly by adding it as a term  $\frac{\alpha \beta}{2} (\frac{1+a}{2}) = \frac{\alpha\beta\gamma}{4N^\theta}$ to the final expression of the free energy \cite{a14} (see Equation (\ref{eq:ARS})).

Now, we can introduce another party made up of $L$ \emph{soft} spins $\left\{z_{\mu}\right\}$, namely i.i.d. variables with an intrinsic standard Gaussian distribution $\mathcal{N}[0,1]$, that interact only with the original party of binary spins $\left\{\sigma_i\right\}$ via the couplings $\left\{\xi_i^{\mu}\right\}$; the related partition function is
\beq\label{eq:Hopfield}
Z_{N,L}(\beta;\xi) =
\sum_{\left\{\sigma\right\}} \int \prod_{\mu} d\mu(z_{\mu})
\;\mbox{exp} \left\{
\sqrt{\frac{\beta}{N^{1-\theta}}}
\sum_{i}^{N} \sum_{\mu}^{L} \xi_i^{\mu}\sigma_i z_{\mu}
\right\}
\:,
\eeq
with $d\mu(z_{\mu})$ standard Gaussian measure for all the $z_{\mu}$. By applying Gaussian integrations as usual \cite{SMnostro}, it is easy to see that $\tilde{Z}_{N,L}(\beta;\xi)$ and $ Z_{N,L}(\beta;\xi)$ are thermodynamically equivalent.
The advantage of the expression (\ref{eq:Hopfield}) is that it is linear  with respect to the memories $\xi_i^{\mu}$, so that the bare topology is simply that of a bipartite random graph with link probability $p_{\mbox{\tiny link}} = (1+a)/2$, like in \cite{agliari2}.

Taken $O$ as a generic observable, depending on the spin configurations $\{ \sigma, z \}$, we define the Boltzmann state $\omega_{\beta}(O)$ at a given value of (fast) noise $\beta$ as
\beq\label{eq_omega}
\omega_{\beta}(O) = Z_{N,L}(\beta;\xi)^{-1}
\sum_{\left\{\sigma\right\}} \int \prod_{\mu} d\mu(z_{\mu})
O(\left\{\sigma,z\right\})
e^{\sqrt{\frac{\beta}{N^{1-\theta}}}\sum_{i}^{N} \sum_{\mu}^{L} \xi_i^{\mu}\sigma_i z_{\mu}},
\eeq
and we introduce a product space on several replicas of the system as
$\Omega_s = \omega_1 \bigotimes \omega_2 \bigotimes ... \bigotimes \omega_s$ \cite{BGG1}.

For a generic function of the memories $F(\xi)$, the quenched average will be defined by the symbol $\mathbb{E}$ and performed in two steps: first we fix the number $l$ of links between the two parties and we perform the average over the Gaussian distribution of the memories:
\beq
\E^{(l)}_{\xi}[F(\xi)] \equiv
\int_{-\infty}^{+\infty} \prod_{(i,\mu)=1}^{l}\frac{d\xi_i^{\mu}}{\sqrt{2\pi}}
e^{-\frac{(\xi_i^{\mu})^2}{2}} F(\xi_i^{\mu}) =
\int F(\xi) \; d\mu_l(\xi)
\equiv f(l)
\:;
\eeq
then, we perform the average over the binomial distribution for the number of links,
\beq
\E_l[f(l)] \equiv \sum_{l=0}^{NL} {NL \choose l} \left(\frac{1+a}{2}\right)^l \left(\frac{1-a}{2}\right)^{(NL-l)}f(l)
\:,
\eeq
so that $\E \equiv \E_l \E^{(l)}_{\xi}$ .
Indeed, for example, $\E[\xi_i^{\mu}]=0$ and $ \E[(\xi_i^{\mu})^2]=(1+a)/2$.

Moreover, as we will see, for a natural introduction of the order parameters, it is useful to define the number of links, $l$, as the product of two independent binomial variables
\beq
l \, \dot{=} \, l_{\eta} l_{\chi}
\:,
\eeq
where the symbol $\dot{=}$ stands for the equality in distribution and
\bea
\label{eq:P(l_eta)}
P(l_{\eta}) &=& {N \choose l_{\eta}} \sqrt{\frac{1+a}{2}}^{l_{\eta}} \sqrt{\frac{1-a}{2}}^{N-l_{\eta}},
\\
P(l_{\chi}) &=& {L \choose l_{\chi}} \sqrt{\frac{1+a}{2}}^{l_{\chi}} \sqrt{\frac{1-a}{2}}^{L-l_{\chi}}
\:.
\eea
Of course, a product of two binomial variables is not a binomial variable itself, so at finite size this definition is not consistent; nevertheless, in the thermodynamic limit, the central limit theorem ensures that only the first two momenta of the distributions survive so that the definitions become consistent.

We also use the symbol $\left\langle \cdot \right\rangle$ to mean  $\left\langle \cdot \right\rangle = \E \Omega (\cdot)$ and $\left\langle \cdot \right\rangle_G = \E^{(l)}_{\xi} \Omega (\cdot)$.

The main thermodynamical quantity of interest is the intensive pressure defined as
\beq\label{eq:pressure}
A(\alpha,\beta) = \lim_{N \to \infty} A_N(\alpha,\beta) = - \beta f(\alpha, \beta) =  \lim_{N \to \infty} \frac{1}{N} \E \log Z_{N,L}(\beta,\xi)
\:.
\eeq
Here $f(\alpha, \beta) = u(\alpha, \beta) - \beta^{-1}s(\alpha, \beta)$ is the free-energy density, $u(\alpha, \beta)$ the internal energy density and $s(\alpha, \beta)$ the entropy density.

Finally, we define two infinite (in the thermodynamic limit) sets of order parameters, the restricted overlaps, as
\bea
q_{12}^{l_{\eta}} = \frac{1}{l_{\eta}} \sum_i^{l_{\eta}} \sigma_i^1 \sigma_i^2,
\nonumber\\
p_{12}^{l_{\chi}} = \frac{1}{l_{\chi}} \sum_{\mu}^{l_{\chi}} z_{\mu}^1 z_{\mu}^2
\:,
\eea
which define the  overlaps (restricted on sub-networks) between two replicas made up by parties with $l_{\eta}$ and $l_{\chi}$ nodes, respectively.

\subsection{Free energy interpolation and general strategy}

In what follows we assume that no real external fields (as magnetic inputs or partial information submission for retrieval) act on the network, but fields insisting on each spin are strictly generated by other spins. Thus, the overall field felt by an element of a given party is the sum (weighted through the couplings), of the states of the spins in  the other party.
Note that spins are connected in loops using the other party as a mirror,
therefore, the equivalent analogue neural network is a recurrent network.

In this section we show that the free-energy can be calculated in specific cases (e.g. at the replica symmetrical level) by using a novel technique that has been developed in \cite{BGG1} for fully connected spin-glass models and extended in \cite{agliari2} to diluted ferromagnetic models.
This technique introduces an external field acting on the system which "imitates" the internal, recurrently-generated input, by reproducing its average statistics.
While the external, fictitious input does not reproduce the statistics of order two and higher, it represents correctly the averages.
These external inputs are denoted as $\eta$ and $\chi$ (one for each spin in each party) and are distributed following the  Gaussian distributions with zero mean and whose variances scale according to the underlying topology (as a function of $\alpha,\theta,\gamma$) and coherently approaches zero when the network topology disappears.

In order to recover the second order statistics, the free-energy is interpolated smoothly between the case in which all fields are external, and all high order statistics is missing, and the case in which all fields are internal, describing the original network: Following the original Guerra's schemes \cite{lele,BGG1,limterm,sumrule}, this allows a powerful sum rule. We use an interpolating parameter $t \in [0,1]$ for this morphing, such that for $t=0$ the fields are all external and the calculation straightforward, while for $t=1$ the original model is fully recovered.

In what follows, for the sake of clearness, we write  $A=A_N(\alpha,\beta)$ even though $\alpha$ should be introduced only once the thermodynamic limit has been performed.
The interpolating quenched pressure $\tilde{A}_N(\alpha,\beta,t)$ at finite $N$ is then defined as
\bea\label{eq:interpol}
&&\!\!\!\!\!\!\!\!\!\!\!\!\!\!\!\!\!\!\!\!\!\!\!\!\!\!\!\!\!\tilde{A}_N(\alpha, \beta; t) =
\frac{1}{N} \E \log \sumc
\exp \left( \sqrt{t} \, \sqrt{\frac{\beta}{N^{1-\theta}}} \sum_{i, \mu}^{N,L} \xi_i^{\mu} \sigma_i z_{\mu} \right)
\cdot
\nonumber\\
&&\!\!\!\!\!\!\!\!\!\!\!\!\!\!\!\!\!\!\!\!\!\!\!\!\!\!\!\!\!\cdot\exp \left( b\,\sqrt{1-t} \sum_{i}^{N} \sigma_i \eta_i \right)
\exp \left( c\,\sqrt{1-t} \sum_{\mu}^{L} z_{\mu} \chi_{\mu} \right)
\exp \left( \frac{d}{2}(1-t) \sum_{\mu}^{L} z_{\mu}^2 \theta_{\mu} \right)
\:.
\eea
Throughout the paper, we assume that the limit $A(\alpha,\beta)=\lim_{N \to \infty}A_N(\alpha,\beta)=\lim_{N \to \infty}A_N(\alpha,\beta,t=1)$ exists.
The ``interpolating fields'' distributions are chosen to mimic the local fields behavior, so that $\eta_i$, $\chi_{\mu}$ and $\theta_{\mu}$ have zero value with probability $\sqrt{(1-a)/2}$, while, with probability $\sqrt{(1+a)/2}$, are normally distributed, except for $\theta_{\mu}$ which assumes value 1\footnote{Indeed, the presence the field $\theta_{\mu}$ has much less physical meaning but simplifies the calculations.}. Consequently, the number of active fields follows Equation  (\ref{eq:P(l_eta)}). As for the constants $b,c,d$, they have to be chosen properly, as shown in the following.

The strategy for the evaluation of the pressure of the original model, $\tilde{A}(\alpha,\beta,t=1)$, is to compute the $t$-streaming of $\tilde{A}(\alpha,\beta,t)$, namely $\partial_t \tilde{A}(\alpha,\beta,t=1)$, and use the fundamental theorem of calculus to obtain
\beq\label{fundamental}
\!\!\!\!\!\!\!\!\!\!\!\!\!\!\!\!\!\!\!\!\!\!\! A_{N}(\alpha,\beta) = \tilde{A}_{N}(\alpha,\beta; t=1) = \tilde{A}_{N}(\alpha,\beta, t=0) + \int_0^1 dt' \left( \frac{d}{dt} \tilde{A}_N(\alpha, \beta; t) \right)_{t=t'}
\:.
\eeq

When evaluating  the streaming $\partial_t \tilde{A}$, we get the sum of four terms ($\mathcal{A}, \mathcal{B}, \mathcal{C}, \mathcal{D}$); each comes as a consequence  of the derivation of a corresponding exponential term appearing into Equation (\ref{eq:interpol}). In order to proceed we need to compute them explicitly:
\bea
\!\!\!\!\!\!\!\!\!\!\!\!\!\!\!\!\!\!\!\!\!\!\!\!\!\!\ \mathcal{A} &=& \frac{1}{N} \E \frac{\sqrt{\beta}}{2\sqrt{t}N^{(1-\theta)/2}}
	\sum_{i,\mu}^{l_\eta, l_\chi} \xi_i^{\mu} \omega(\sigma_i,z_{\mu}) = \frac{1}{N} \E \frac{\sqrt{\beta}}{2\sqrt{t}N^{(1-\theta)/2}}
	\sum_{i,\mu}^{l_\eta, l_\chi} \partial_{\xi_i^{\mu}} \omega(\sigma_i,z_{\mu})
\nonumber\\
\!\!\!\!\!\!\!\!\!\!\!\!\!\!\!\!\!\!\!\!\!\!\!\!\!\!\ \phantom{A} &=&
\frac{1}{N} \E \frac{\beta}{2N^{(1-\theta)}}
	\sum_{i,\mu}^{l_\eta, l_\chi} \left[ \omega(\sigma_i^2,z_{\mu}^2) - \omega^2(\sigma_i,z_{\mu}) \right] =
	\nonumber\\
\!\!\!\!\!\!\!\!\!\!\!\!\!\!\!\!\!\!\!\!\!\!\!\!\!\!\ \phantom{A} &=&
\frac{1}{N} \frac{\beta}{2N^{(1-\theta)}}
	\sum_{l_{\eta},l_{\chi}} P(l_{\eta}, l_{\chi}) l_{\eta} l_{\chi}
	\left[ \langle z_{\mu}^2\rangle_G - \langle q_{12}^{l_{\eta}}  p_{12}^{l_{\chi}}\rangle_G \right] =
\nonumber\\
\!\!\!\!\!\!\!\!\!\!\!\!\!\!\!\!\!\!\!\!\!\!\!\!\!\!\ \phantom{A} &=&
\frac{1}{N} \frac{\beta}{2N^{(1-\theta)}}
	NL\ap
	\left[ \langle z_{\mu}^2\rangle - \langle q_{12}^{l_{\eta}}  p_{12}^{l_{\chi}}\rangle \right] =\frac{\alpha\beta}{2} \frac{\gamma}{2}
	\left[ \langle z_{\mu}^2\rangle - \langle q_{12}^{l_{\eta}} p_{12}^{l_{\chi}}\rangle \right]
\:,
\eea
where in the first passage we used integration by parts and, in the fourth, the factorization properties of the quenched averages \cite{agliari2,AC,barra4,contu,GG} (which should be understood in the thermodynamic limit).

The same procedure can be used in the computation of the other terms, so to get:
\bea
\!\!\!\!\!\!\!\!\!\!\!\!\!\!\!\!\!\!\!\!\!\!\!\!\!\!\  \mathcal{B} &=&
-\frac{1}{N} \E \frac{b}{2\sqrt{1-t}}
	\sum_{i}^{l_\eta} \eta_i \omega(\sigma_i) = -\frac{1}{N} \E \frac{b}{2\sqrt{1-t}}
	\sum_{i}^{l_\eta} \partial_{\eta_i} \omega(\sigma_i)
\nonumber\\
\!\!\!\!\!\!\!\!\!\!\!\!\!\!\!\!\!\!\!\!\!\!\!\!\!\!\  \phantom{B} &=&
-\frac{1}{N} \E \frac{b^2}{2}
	\left( l_{\eta} - \sum_{i}^{l_\eta} \omega^2(\sigma_i) \right)=-\frac{b^2}{2N}
	\sum_{l_{\eta}} P(l_{\eta})\, l_{\eta} \left( 1 -  \langle q_{12}^{l_{\eta}} \rangle_G \right)
\nonumber\\
\!\!\!\!\!\!\!\!\!\!\!\!\!\!\!\!\!\!\!\!\!\!\!\!\!\!\  \phantom{B} &=&
- \frac{b^2}{2}
	\ap^{\frac{1}{2}} \left( 1 - \langle q_{12}^{l_{\eta}} \rangle \right)- \frac{b^2}{2}
	\sqrt{\frac{\gamma}{2}} N^{-\frac{\theta}{2}} \left( 1 - \langle q_{12}^{l_{\eta}} \rangle \right)
\:;
\eea
\bea
\!\!\!\!\!\!\!\!\!\!\!\!\!\!\!\!\!\!\!\!\!\!\!\!\!\!\  \mathcal{C} &=& -\frac{1}{N} \E \frac{c}{2\sqrt{1-t}}
	\sum_{\mu}^{l_\chi} \chi_{\mu} \omega(z_{\mu}) =-\frac{1}{N} \E \frac{c}{2\sqrt{1-t}}
	\sum_{\mu}^{l_\chi} \partial_{\chi_\mu} \omega(z_{\mu}) =
\nonumber\\
\!\!\!\!\!\!\!\!\!\!\!\!\!\!\!\!\!\!\!\!\!\!\!\!\!\!\  \phantom{C} &=& -\frac{1}{N} \E \frac{c^2}{2}
	\left( \sum_{\mu}^{l_\chi} \omega(z_{\mu}^2) - \sum_{\mu}^{l_\chi} \omega^2(z_{\mu}) \right)=-\frac{c^2}{2N}
	\sum_{l_{\chi}} P(l_{\chi})\, l_{\chi} \left(\sum_{\mu}^{l_\chi} \langle z_{\mu}^2\rangle_G -  \langle p_{12}^{l_{\chi}}\rangle_G \right)
\nonumber\\
  \phantom{C} &=& - \frac{\alpha c^2}{2}
	\ap^{\frac{1}{2}} \left( \langle z^2 \rangle - \langle p_{12}^{l_{\chi}} \rangle \right) =
	- \frac{\alpha c^2}{2}
	\sqrt{\frac{\gamma}{2}} N^{-\frac{\theta}{2}} \left( \langle z^2 \rangle - \langle p_{12}^{l_{\chi}} \rangle \right)
\:;
\eea
\bea
\!\!\!\!\!\!\!\!\!\!\!\!\!\!\!\!\!\!\!\!\!\!\!\!\!\!\ 
\mathcal{D} &=& -\frac{1}{N} \E \frac{d}{2}
	\sum_{\mu}^{l_\chi} \omega(z_{\mu}^2) = -\frac{\alpha d}{2}
	\ap^{\frac{1}{2}} \langle z^2 \rangle
\:.
\eea

Now, the $t$-streaming of the pressure reads off as
\bea\label{streaming}
\!\!\!\!\!\!\!\!\!\!\!\!\!\!\!\!\!\!\!\!\!\!\!\!\!\!\!\!\!\!\!\!\frac{d \tilde{A}_N(\alpha,\beta,t)}{d t} &=& \left[ \frac{\alpha\beta}{2}N^{\theta} \ap - \frac{\alpha c^2}{2}\ap^{\frac{1}{2}} - \frac{\alpha d}{2}\ap^{\frac{1}{2}} \right] \langle z^2 \rangle +
\nonumber\\
\!\!\!\!\!\!\!\!\!\!\!\!\!\!\!\!\!\!\!\!\!\!\!\!\!\!\!\!\!\!\!\!\phantom{A}&\phantom{=}&
- \frac{\alpha\beta}{2} N^{\theta} \ap \langle q_{12}^{l_{\eta}} p_{12}^{l_{\chi}} \rangle
+ \frac{b^2}{2}\ap^{\frac{1}{2}} \langle q_{12}^{l_{\eta}} \rangle
+ \frac{c^2}{2}\ap^{\frac{1}{2}} \langle p_{12}^{l_{\chi}} \rangle +
\nonumber\\
\!\!\!\!\!\!\!\!\!\!\!\!\!\!\!\!\!\!\!\!\!\!\!\!\!\!\!\!\!\!\!\!&\phantom{=}&-
\ap^{\frac{1}{2}} \frac{b^2}{2}
\:.
\eea

\subsection{Replica symmetric approximation and fluctuation source}

As it is, this streaming encodes the whole full replica-symmetry-breaking  complexity \cite{MPV,nishimori} of the underlying glassy phase and it is intractable.
Our plan is to split this derivative in two terms, one dealing with the averages of the order parameters and one accounting for their fluctuations. To this aim we introduce the \emph{source of fluctuations}, $S(\alpha,\beta,t)$, as
\beq\label{kingkong}
S(\alpha,\beta,t) = - \frac{\alpha \beta}{2} \ap N^{\theta} \langle (q_{12}^{l_{\eta}}-\overline{q})(p_{12}^{l_{\chi}}- \overline{p}) \rangle,
\eeq
with
\beq
\overline{q} \, \equiv \, \E_{l_{\eta}} \overline{q}^{l_{\eta}}, \ \ \overline{p} \, \equiv \, \E_{l_{\eta}} \overline{p}^{l_{\eta}}
\:.
\eeq
Notice that the main order parameters $\overline{q}$ and $\overline{p}$ sum every overlap, each with its relative weight, on every possible subnetwork of the whole network according to the approaches \cite{ton1,ton2} and that they recover the standard order parameters of the Hopfield model when dilution is neglected \cite{peter,BGG1}.

In order to relate Equation (\ref{kingkong}) to Equation (\ref{streaming}), let us remember that we still have free parameters that can be chosen as \footnote{In particular, we choose $d$ to cancel the $\langle z^2 \rangle$ terms appearing in the first line of equation (\ref{streaming}).}
\bea
b &=& \sqrt{\alpha\beta \ap^\frac{1}{2} N^\theta \overline{p}} =
\sqrt{\alpha\beta\overline{p}} \left(\frac{\gamma}{2}\right)^{\frac{1}{4}}N^{\frac{\theta}{4}},
\nonumber\\
c &=& \sqrt{\beta \ap^{\frac{1}{2}} N^\theta \overline{q}} =
\sqrt{\beta\overline{q}} \left(\frac{\gamma}{2}\right)^{\frac{1}{4}} N^{\frac{\theta}{4}},
\nonumber\\
d &=& \beta N^{\theta} \ap^{\frac{1}{2}} (1- \overline{q}) =
\beta \left(\frac{\gamma}{2}\right)^{\frac{1}{2}} N^{\frac{\theta}{2}}(1-\overline{q})
\:,
\eea
so to get
\beq\label{dAdt}
  \frac{d \tilde{A}(\alpha,\beta;t)}{dt} = S(\alpha,\beta,t) +\frac{\alpha \beta}{2} \ap N^{\theta}\, \overline{p} (\overline{q}-1) =
S(\alpha,\beta,t) + \frac{\alpha \beta \gamma}{4} \overline{p} (\overline{q} -1)
\:.
\eeq

In the replica symmetric approximation, the order parameters do not fluctuate with respect to their quenched average in the thermodynamic limit as they get delta-distributed over their replica symmetric averages $\overline{q}, \ \overline{p}$, which have been denoted with a bar.
As a consequence, within this approximation, we can neglect the fluctuation source term $S(\alpha,\beta,t)$ and keep only the replica symmetric overlap averages in the expression (\ref{dAdt}) such that its integration is trivially reduced to a multiplication by one.

In order to obtain an explicit expression of the sum rule (\ref{fundamental}), we can then proceed to analyze the starting point for the ''morphing'', namely $\tilde{A}(\alpha,\beta;t=0)$, which can be calculated straightforwardly as it involves only one-body interactions:
\bea\label{eq:A0}
  &&\tilde{A}_N(\alpha,\beta,t=0) = \\ \nonumber
  &&=\frac{1}{N} \E \log \sumc
\exp \bigg (~b\, \sum_{i}^{N} \sigma_i \eta_i + c\, \sum_{\mu}^{L} z_{\mu} \chi_{\mu}+ \frac{d}{2}\sum_{\mu}^{L} \theta_\mu z_{\mu}^2 \bigg)=\\
  &&=  \sqrt{\frac{\gamma}{2}}N^{-\frac{\theta}{2}}\log 2 +  \sqrt{\frac{\gamma}{2}}N^{-\frac{\theta}{2}}\int d\mu(\eta) \log \mathrm{cosh}(\sqrt{\alpha\beta\overline{p}} \left(\frac{\gamma}{2}\right)^{\frac{1}{4}}N^{\frac{\theta}{4}}~ \eta) +
\nonumber\\
  &&+\frac{\gamma \alpha \beta \bar{q}}{4} \frac{1}{1- \beta \left(\frac{\gamma}{2}\right)^{\frac{1}{2}} N^{\frac{\theta}{2}}(1-\overline{q}) }-\frac{\alpha}{2} \left(\frac{\gamma}{2}\right)^{\frac{1}{2}}  N^{-\frac{\theta}{2}} \log \left( 1- \beta \left(\frac{\gamma}{2}\right)^{\frac{1}{2}} N^{\frac{\theta}{2}}(1-\overline{q})  \right), \nonumber
\eea
where we used
\bea\label{eq:A02}
  &&\frac{1}{N} \E_{l_\eta} \E^{l_\eta}_{\eta} \log \sum_{\{\sigma\}} \exp \left( b\, \sum_{i}^{N} \sigma_i \eta_i \right) = \nonumber \\
  &&=\frac{1}{N} \E_{l_\eta} \E^{l_\eta}_{\eta} \sum_{i}^{N} \log 2\mathrm{cosh}(b~ \eta_i) =
\frac{1}{N} \E_{l_\eta} l_\eta ~\E_{\eta}  \log 2\mathrm{cosh}(b~ \eta)= \nonumber \\
  &&=  \ap^{\frac{1}{2}}\log 2 +  \ap^{\frac{1}{2}}  \int d\mu(\eta) \log \mathrm{cosh}(b~ \eta),
\eea
and
\bea\label{eq:A03}
  && \frac{1}{N} \E_{l_\chi} \E^{l_\chi}_{\chi} \log \int \prod_{\mu} d\mu(z_\mu) \exp \left( c\, \sum_{\mu}^{L} z_{\mu} \chi_{\mu} \right) \exp \left( \frac{d}{2}\sum_{\mu}^{L} \theta_\mu z_{\mu}^2\right)= \nonumber\\
  && =\frac{1}{N} \E_{l_\chi} \E^{l_\chi}_{\chi}  \left(\sum_{\mu} \frac{c^2  \chi_\mu^2}{2 (1-d \theta_\mu)}\right)+ \frac{1}{N} \E_{l_\chi} \E^{l_\chi}_{\chi}\frac{1}{2} \sum_{\mu}  \log \left( \frac{1}{1-d \theta_\mu} \right)=\nonumber\\
  && =\frac{\alpha}{2}  \sqrt{\ap} \left( \frac{c^2  \sigma_\chi^2}{(1-d)}\right)-  \frac{\alpha}{2}  \sqrt{\ap} \log \left( 1-d \right)=\nonumber\\
  &&=\frac{\gamma \alpha \beta \bar{q}}{4} \frac{1}{1- \beta \left(\frac{\gamma}{2}\right)^{\frac{1}{2}}N^{\frac{\theta}{2}} (1-\overline{q})}-\frac{\alpha}{2} \left(\frac{\gamma}{2}\right)^{\frac{1}{2}}  N^{-\frac{\theta}{2}} \log \left( 1- \beta \left(\frac{\gamma}{2}\right)^{\frac{1}{2}} N^{\frac{\theta}{2}}(1-\overline{q})  \right).
\eea

Now, substituting the expression for $\tilde{A}_N(\alpha,\beta,t=0)$ of Equation(\ref{eq:A0}) into (\ref{fundamental}), we obtain the replica-symmetric free energy (strictly speaking the mathematical pressure) of the network as
\bea\label{eq:ARS}
\!\!\!\!\!\!\!\!\!\!\!\!\!\!\!\!\!\! &&\tilde{A}_N^{RS}(\alpha,\beta) =\tilde{A}_N(\alpha,\beta,t=0)+\left.\frac{d \tilde{A}_N^{RS}(\alpha,\beta,t}{dt}\right|_{t=0}= \nonumber \\
\!\!\!\!\!\!\!\!\!\!\!\!\!\!\!\!\!\! &&=\sqrt{\frac{\gamma}{2}}N^{-\frac{\theta}{2}}\left[\log 2 +  \int d\mu(\eta) \log \mathrm{cosh}\left(\sqrt{\alpha\beta\overline{p}} \left(\frac{\gamma}{2}\right)^{\frac{1}{4}}N^{\frac{\theta}{4}}
~ \eta\right)\right]+
\nonumber\\
\!\!\!\!\!\!\!\!\!\!\!\!\!\!\!\!\!\! &&+\frac{\alpha \beta \gamma \ \bar{q}}{4} \frac{1}{1- \beta \left(\frac{\gamma}{2}\right)^{\frac{1}{2}} N^{\frac{\theta}{2}}(1-\overline{q}) }
-\frac{\alpha}{2} \left(\frac{\gamma}{2}\right)^{\frac{1}{2}}  N^{-\frac{\theta}{2}} \log \left( 1- \beta \left(\frac{\gamma}{2}\right)^{\frac{1}{2}} N^{\frac{\theta}{2}}(1-\overline{q})  \right) +
\nonumber\\
\!\!\!\!\!\!\!\!\!\!\!\!\!\!\!\!\!\! && +\frac{\alpha \beta \gamma}{4} \overline{p} (\overline{q} -1)+\frac{\alpha\beta\gamma}{4}N^{-\theta}
\:.
\eea
Despite the last expression is meant to hold in the thermodynamic limit, with a little mathematical abuse we left the explicit dependence on $N$ to discuss some features of the solution: 
Equation (\ref{eq:ARS}) may look strange due to the strong presence of various powers of the volume size $N$, which in principle are potentially unwanted divergencies. We start noticing that, in the limit of zero dilution $\theta = 0$ and homogeneous distribution of fields $\gamma=2$, the expression for the free-energy recovers the replica symmetric one of the analogue Hopfield model \cite{BGG1} (or digital one without retrieval \cite{amit}). Moreover, remembering the various topological regimes outlined in Section $3$, we see that when the network changes the topological phase, for instance moving from a fully connected topology to a sparse graph, the coordination number may scale with the volume size or remain constant. These situations are deeply different from a thermodynamical viewpoint because, in order to have no negligible contributions to the free-energy, fields obtained by an extensive number of (finite) terms in the fully connected scenario must be (possibly) turned into fields obtained by a finite number of (infinite) terms in the dilute regime. As the topology changes, the fields must follow accordingly, which is equivalent to a (fast) noise rescaling with the volume size that is another standard approach to diluted network \cite{agliari1,barra4,gt2}.

The physical free-energy is then obtained by extremizing this expression with respect to the order parameters; we only stress here that, as a general property of these  neural networks/bipartite spin-glasses, the free-energy now obeys a min-max principle, which will not be deepened here (because it does not change the following procedure and it has been discussed in \cite{BGG1}). As a consequence, the following system defines the values of the overlaps (as functions of $\alpha, \ \beta$) that must be used into Equation (\ref{eq:ARS})
\bea
\frac{\partial \tilde{A}}{\partial \overline{q}} &=&
\frac{\alpha\beta(\frac{\gamma}{2})}{2}
\left( \overline{p} - \frac{(\frac{\gamma}{2})^{\frac{1}{2}} N^{\frac{\theta}{2}}\beta \overline{q}}{(1-\beta(\frac{\gamma}{2})^{\frac{1}{2}}N^{\frac{\theta}{2}}(1-\overline{q}))^2} \right) = 0
\nonumber\\
\frac{\partial \tilde{A}}{\partial \overline{p}} &=&
\frac{\alpha\beta(\frac{\gamma}{2})}{2}
\left[ \overline{q} - \int d\mu({\eta})
\,\mbox{tanh}^2(\sqrt{\alpha\beta\overline{p}} \left(\frac{\gamma}{2}\right)^{\frac{1}{4}} N^{\frac{\theta}{4}} \eta)\right]
=0
\:,
\eea
by which
\beq\label{eq:selfcons}
\overline{q} =
\int d\mu({\eta})
\,\mbox{tanh}^2 \left( \frac{\sqrt{\alpha \overline{q}} \beta ({\frac{\gamma}{2}})^{\frac{1}{2}} N^{\frac{\theta}{2}}}{1-\beta(\frac{\gamma}{2})^{\frac{1}{2}}N^{\frac{\theta}{2}}(1-\overline{q})}\, \eta\right)
\:.
\eeq

All the related models (e.g. Viana-Bray \cite{viana}, Hopfield \cite{hopfield}, Sherrigton-Kirkpatrick \cite{david}) display an ergodicity breaking associated with a second order phase transition and presence of criticality. If we assume the same behavior even for the model investigated here, the self-consistency equation (\ref{eq:selfcons}) can give hints on the critical line (in the parameter space) where ergodicity breaks down. In fact, when leaving the ergodic region (implicitly defined by $\overline{q}=0, \ \overline{p}=0$) the order parameters start  growing (implicitly defining the critical line as the starting point)  and, as continuity is assumed through the second order kind of transition, we can expand the r. h. s. of Equation (\ref{eq:selfcons}) for low $\overline{q}$ and obtain a polynomial expression on both sides. Then, due to the principle of identity of polynomials, we can equate the two sides term by term obtaining
\beq\label{critica1}
\beta_c = \frac{1}{\left(\frac{\gamma}{2}\right)^{\frac{1}{2}} N^{\frac{\theta}{2}} \left( 1 + \sqrt{\alpha} \right)}
\:,
\eeq
which is the critical surface of the system.

Mirroring the discussion dealing with the free energy, we note that this result too is clearly a consequence of the choice (\ref{eq:normalization}) for the normalization factor that gives us an extensive thermodynamics. If we normalize choosing $D=N$, as it is usual in the Hopfield model \cite{amit}, we obtain (turning to $T= 1/\beta$ which is most intuitive) $T_c = N^{\theta/2}(1+\sqrt{\alpha})\sqrt{\gamma/2}$ (and recover the AGS line for $\theta=0$ and $\gamma=2$), such that the overall effect of increasing dilution is to reduce the value of the critical temperature because the couplings, on average, become weaker. In particular, in the finite connectivity regime $(\theta=1)$, the network is built of by $N$ links instead of $N^2$ which, roughly speaking, implies a rescaling in the temperature proportional to $\sqrt{N}$ (coherently with a spin-glass behavior), as in the ferromagnetic counterpart its rescale is ruled by $N$ instead of $\sqrt{N}$ \cite{agliari1} because the latter is a model defined through the first momentum, while the former by the variance.

Furthermore we stress that the system displays only one critical surface splitting the ergodic region from the spin glass and there are no further `weak-transitions" for each sub-overlap, coherently with the scenario discussed in \cite{BarraMO} for the similar case of the Viana-Bray model \cite{viana}.

\section{Fluctuation theory and critical behavior}

The plan of this section is studying the regularity of the rescaled (and centered) overlap correlation functions.
\newline
The idea is as follows: If the system undergoes a second order phase transition, the (extensive) fluctuations of its order parameters should diverge on the critical surface (\ref{critica1}), hence they should be described by meromorphic functions; from the poles of these functions it is possible to detect the critical surface. As a consequence, an explicit knowledge of these functions would confirm (or reject) the critical picture we obtained through the small overlap expansion of the previous section. However, obtaining them explicitly is not immediate and we sketch in what follows our strategy.
At first, we define the (rescaled and centered) fluctuations of the order parameters as
\bea
Q_{ab}^{l_{\eta}} &=& \sqrt{N} \left( q_{ab}^{l_{\eta}} - \overline{q}^{l_{\eta}} \right),
\nonumber\\
P_{ab}^{l_{\eta}} &=& \sqrt{L} \left( p_{ab}^{l_{\chi}} - \overline{p}^{l_{\eta}} \right),
\eea
such that, while $q_{ab}^{l_{\eta}} \in [-1,+1], \ p_{ab}^{l_{\eta}} \in [-1,+1]$, $Q_{ab}^{l_{\eta}} \in \mathbb{R}, \ P_{ab}^{l_{\eta}} \in \mathbb{R}$, hence, the square of the latter may diverge as expected for second order phase transitions.

Nevertheless, obtaining them explicitly from the original Hamiltonian is prohibitive and we use another procedure, originally outlined in \cite{sumrule}: We evaluate these rescaled overlap fluctuations weighted with the non-interacting Hamiltonian in the Maxwell-Boltzman distribution, hence $\langle
Q_{l_\eta 12}^2 \rangle_{t=0}$, $\langle Q_{l_\eta 12} P_{l_\chi 12}
\rangle_{t=0}$, $\langle P_{l_\chi 12}^2 \rangle_{t=0}$, then we derive
the streaming of a generic observable $O$ (that is in principle a function of the spins of the parties and of the quenched memories), namely $\partial_t \langle O \rangle_t$ such that we know how to propagate $\langle O \rangle_{(t=0)}$ up to $\langle O \rangle_{(t=1)}$ (which is our goal), and finally we use this streaming equation (which turns out to be a dynamical system) on the Cauchy problem defined by $\langle
Q_{l_\eta 12}^2 \rangle_{t=0}$, $\langle Q_{l_\eta 12} P_{l_\chi 12}
\rangle_{t=0}$, $\langle P_{l_\chi 12}^2 \rangle_{t=0}$, obtaining the attended result. Once the procedure is completed, the simple analysis of the poles of $\langle
Q_{l_\eta 12}^2 \rangle_{t=1}$, $\langle Q_{l_\eta 12} P_{l_\chi 12}
\rangle_{t=1}$, $\langle P_{l_\chi 12}^2 \rangle_{t=1}$ will identify the critical surfaces of the system.

Starting with the study of the structure of the derivative, our aim is to compute the $t$-streaming for a generic observable $O_s$ of $s$ replicas.
Calling
\bea
\!\!\!\!\!\!\!\!\!\!\!\!\!\!\!\!\!\!\!\!\!
H_s &=&\sum_{a=1}^s \left\{ \sqrt{t} \sqrt{\frac{\beta}{N^{1-\theta}}} \sum_{i,\mu}^{l_{\eta},l_{\chi}} \xi_i^{\mu} \sigma_i^a z_{\mu}^a + \sqrt{1-t} ~b \sum_i^{l_{\eta}} \eta_i \sigma_i^a + \sqrt{1-t} ~c \sum_{\mu}^{l_{\chi}} \chi_{\mu} z_{\mu}^a + \right.
\nonumber\\
\!\!\!\!\!\!\!\!\!\!\!\!\!\!\!\!\!\!\!\!\!&\phantom{=}&
\left.+\,(1-t)\frac{d}{2} \theta_{\mu} \sum_{\mu}^{l_{\chi}} (z_{\mu}^a)^2 \right\}
\:,
\eea
such that
$$
\langle O \rangle_t = \frac{\int \prod_{\mu}^L d\mu(z_{\mu}) \sum_{\sigma} O \exp\left( -\beta H_s \right)} {\int \prod_{\mu}^L d\mu(z_{\mu}) \sum_{\sigma} \exp\left( -\beta H_s \right)},
$$
its  $t$-streaming is
\beq
  \frac{d \<O_s\>_t }{d t} =
\frac{d}{dt} \E \frac{\sumc O_s e^{H_s}}{\sumc e^{H_s}} =
\E \left[\Omega \left(O_s \frac{dH}{dt} \right)\right] - \E  \left[\Omega(O_s) \Omega \left( \frac{dH}{dt} \right)\right]
\:.
\eeq
In the last equation eight terms contribute. Let us call them $A_1$,  $B_1$,  $C_1$,  $D_1$, $A_2$,  $B_2$,  $C_2$,  $D_2$ and compute them explicitly:
\bea
  A_1 &=&
\frac{1}{2} \sqrt\frac{\beta}{t N^{1-\theta}} \E \sumc O_s \sum_a  \sum_{i,\mu}^{l_{\eta},l_{\chi}}
\xi_i^{\mu} \sigma_i^a z_{\mu}^a \frac{e^{H_s}}{Z_s}=
\nonumber\\
  \phantom{A} &=&
\frac{1}{2} \sqrt\frac{\beta}{t N^{1-\theta}} \E \sumc O_s \sum_a  \sum_{i,\mu}^{l_{\eta},l_{\chi}}
\sigma_i^a z_{\mu}^a  \partial_{\xi_i^{\mu}} \frac{e^{H_s}}{Z_s} =
\nonumber\\
 \phantom{A} &=&
\frac{\beta}{2N^{1-\theta}} \E \left[ \sum_{a,b}^s \sumimu \Omega(O_s \sigma_i^a z_{\mu}^a \sigma_i^b z_{\mu}^b) -  \sum_{a,b}^s \sumimu \Omega(O_s \sigma_i^a z_{\mu}^a) \Omega( \sigma_i^b z_{\mu}^b) \right] =
\nonumber\\
  &=&
\frac{\beta}{2N^{1-\theta}} \E
\left[ \sum_a l_{\eta} \summu \Omega(O_s (z^a)^2) + \sum_{a\neq b}^s l_{\eta} l_{\chi}  \Omega(O_s q_{ab}^{l_{\eta}} p_{ab}^{l_{\chi}}) -  s \, l_{\eta}l_{\chi} \sum_{a}\Omega(O_s q_{s+1,a}^{l_{\eta}} p_{s+1,a}^{l_{\chi}})\right]=
\nonumber\\
 \phantom{A} &=&
\frac{\beta L N^{\theta} }{2} \left( \frac{1+a}{2} \right)
\left[ \sum_a \langle O_s (z^a)^2\rangle + \sum_{a\neq b}^s \langle O_s q_{ab}^{l_{\eta}} p_{ab}^{l_{\chi}} \rangle -  s \sum_{a} \langle O_s q_{s+1,a}^{l_{\eta}} p_{s+1,a}^{l_{\chi}} \rangle \right]
\:.
\eea
\bea
  A_2 &=&
-\frac{1}{2} \sqrt\frac{\beta}{t N^{1-\theta}} \E \Omega{O_s} \sumc \sum_a  \sum_{i,\mu}^{l_{\eta},l_{\chi}}
\xi_i^{\mu} \sigma_i z_{\mu} \frac{e^{H_s}}{Z_s}=
\nonumber\\
  \phantom{A} &=&
-\frac{1}{2} \sqrt\frac{\beta}{t N^{1-\theta}} \E \Omega{O_s} \sumc \sum_a  \sum_{i,\mu}^{l_{\eta},l_{\chi}}
\sigma_i z_{\mu}  \partial_{\xi_i^{\mu}} \frac{e^{H_s}}{Z_s} =
\nonumber\\
  \phantom{A} &=&
-\frac{\beta}{2N^{1-\theta}} \E \Omega{O_s} \left[ \sum_{a,b}^s \sumimu \Omega( \sigma_i^a z_{\mu}^a \sigma_i^b z_{\mu}^b) -  \sum_{a,b}^s \sumimu \Omega( \sigma_i^a z_{\mu}^a) \Omega( \sigma_i^b z_{\mu}^b) \right] =
\nonumber\\
  \phantom{A} &=&
-\frac{\beta}{2N^{1-\theta}} \E
\left[ \sum_{a<b} \summu \Omega(O_s \sigma_i^{s+1} z_{\mu}^{s+1} \sigma_i^{s+2} z_{\mu}^{s+2}) + \sum_{a}^s  \Omega(O_s) \Omega((z_{\mu}^a)^2) \,+  \right.
\nonumber\\
  \phantom{A} &\phantom{=}&
\left. + \phantom{\frac{A}{B}} s^2 \Omega(O_s \sigma_i^{s+1} z_{\mu}^{s+1} \sigma_i^{s+2} z_{\mu}^{s+2}) \right]=
\nonumber\\
  \phantom{A} &=&
-\frac{\beta}{2N^{1-\theta}} \E
\left[ \frac{s(s-1)}{2} \, l_{\eta} l_{\chi} \Omega(O_s q_{s+1,s+2}^{l_{\eta}} p_{s+1,s+2}^{l_{\chi}})  + s l_{\eta} l_{\chi} \Omega(O_s (z_{\mu}^{s+1})^2) + \right.
\nonumber\\
  \phantom{A} &\phantom{=}&
\left.
-  \,s^2 \Omega(O_s q_{s+1,s+2}^{l_{\eta}} p_{s+1,s+2}^{l_{\chi}}) \right]=
\nonumber\\
  \phantom{A} &=&
-\frac{\beta L N^{\theta}}{2} \ap
\left[ s \langle O_s q_{s+1,s+2}^{l_{\eta}} p_{s+1,s+2}^{l_{\chi}}\rangle - s  \langle O_s (z^{s+1})^2 \rangle \right]
\:.
\eea
With analogous calculations
\bea
\!\!\!\!\!\!\!\!\!\!\!\!\!\!\!\!\!\!\!\!\!B_1 &=&
-\frac{b^2}{2} \ap^{\frac{1}{2}} N
\left[ \sum_{a<b} \langle O_s q_{ab}^{l_{\eta}} \rangle - s  \langle O_s q_{s+1 a}^{l_{\eta}} \rangle
+s  \langle O_s  \rangle \right]
\:,
\\
\!\!\!\!\!\!\!\!\!\!\!\!\!\!\!\!\!\!\!\!\!B_2 &=&
\frac{b^2}{2} \ap^{\frac{1}{2}} N
\left[ -s \langle O_s q_{s+1,s+2}^{l_{\eta}} \rangle + s  \langle O_s  \rangle
\right]
\:,
\eea
\bea
\!\!\!\!\!\!\!\!\!\!\!\!\!\!\!\!\!\!\!\!\!C_1 &=&
-\frac{c^2}{2} \alpha \ap^{\frac{1}{2}} N
\left[ \sum_a \langle O_s  ((z^a)^2) \rangle + \sum_{a<b} \langle O_s p_{a,b}^{l_{\chi}} \rangle - s \sum_a \langle O_s p_{s+1,a}^{l_{\chi}} \rangle
\right]
\:,
\\
\!\!\!\!\!\!\!\!\!\!\!\!\!\!\!\!\!\!\!\!\!C_2 &=&
-\frac{c^2}{2} \alpha \ap^{\frac{1}{2}} N
\left[ s \langle O_s  (z^{s+1})^2 \rangle - \frac{s(s+1)}{2} \,  \langle O_s p_{s+1,s+2}^{l_{\chi}} \rangle
\right]
\:,
\eea
\bea
\!\!\!\!\!\!\!\!\!\!\!\!\!\!\!\!\!\!\!\!\!D_1 &=&
-\frac{d\alpha N}{2} \ap^{\frac{1}{2}} \sum_a \langle O_s (z^a)^2 \rangle
\:,
\\
\!\!\!\!\!\!\!\!\!\!\!\!\!\!\!\!\!\!\!\!\!D_2 &=&
\frac{d\alpha N}{2} \ap^{\frac{1}{2}} s \langle O_s (z^{s+1})^2 \rangle
\:.
\eea

Therefore, merging all these terms together, the streaming is
\bea
\frac{d}{dt} \langle O_s \rangle_t  &=&
\beta \sqrt{\alpha} \frac{\gamma}{2} \left( \sum_{a<b}^s \langle O Q_{ab}^{l_{\eta}} P_{ab}^{l_{\chi}} \rangle_t
- s \sum_{a}^s \langle O Q_{a, s+1}^{l_{\eta}} P_{a, s+1}^{l_{\chi}} \rangle_t  \,+ \right.
\nonumber\\
&\phantom{=}&\left. +\,\frac{s(s+1)}{2} \langle O Q_{s+1, s+2}^{l_{\eta}} P_{s+1,s+2}^{l_{\chi}} \rangle_t \right).
\eea

\vspace{1pc}

In order to control the overlap fluctuations, namely $\langle
Q_{l_\eta 12}^2 \rangle_{t=1}$, $\langle Q_{l_\eta 12} P_{l_\chi 12}
\rangle_{t=1}$, $\langle P_{l_\chi 12}^2 \rangle_{t=1}$,  ..., noting that
the streaming equation pastes two replicas to the ones already
involved ($s=2$ so far), we need to study nine correlation
functions. It is  then useful to introduce them and refer to them by
capital letters so to simplify their visualization:

\begin{eqnarray}
\langle Q_{l_\eta12}^2 \rangle_t &=& A(t), \ \ \  \langle
Q_{l_\eta12}Q_{l_\eta 13}
\rangle_t = B(t), \ \ \ \langle Q_{l_\eta12}Q_{l_\eta 34}\rangle_t = C(t), \\
\langle Q_{l_\eta 12}P_{l_\chi12} \rangle_t &=& D(t), \ \ \ \langle
Q_{l_\eta12}P_{l_\chi 13} \rangle_t = E(t), \ \ \ \langle
Q_{l_\eta12}P_{l_\chi 34}\rangle_t = F(t), \\
\langle P_{l_\chi 12}^2 \rangle_t &=& G(t), \ \ \ \langle
P_{l_\chi 12}P_{l_\chi 13} \rangle_t = H(t), \ \ \ \langle
P_{l_\chi 12}P_{l_\chi 34}\rangle_t = I(t).
\end{eqnarray}

Let us now sketch their streaming. First, we introduce the
operator ``dot'' as
$$
\dot{O}  = \frac{2}{\beta\sqrt{\alpha}\gamma}\frac{d O}{dt},
$$
which simplifies calculations and shifts the propagation of the
streaming from $t=1$ to $t=\beta\sqrt{\alpha} \gamma/ 2$. Using this we sketch
how to write the streaming of the first two correlations (as it
works in the same way for any other):
\begin{eqnarray}
  \dot{A} &=& \langle Q_{l_\eta 12}^2 Q_{l_\eta 12}P_{l_\chi12} \rangle_t -4
\langle Q_{l_\eta 12}^2Q_{l_\eta 13}P_{l_\chi 13} \rangle_t + 3 \langle Q_{l_\eta 12}^2 Q_{l_\eta 34}P_{l_\chi 34} \rangle_t, \nonumber\\    \dot{B} &=&\langle Q_{l_\eta 12}Q_{l_\eta 13}\Big( Q_{l_\eta 12}P_{l_\chi 12} +Q_{l_\eta 13}P_{l_\chi 13} + Q_{l_\eta 23}P_{l_\chi 23} \Big)\rangle_t - \nonumber\\
  &-& 3\langle Q_{l_\eta 12}P_{l_\chi 13} \Big( Q_{l_\eta 14}P_{l_\chi 14} +
Q_{l_\eta 24}P_{l_\chi 24} + Q_{l_\eta 34}P_{l_\chi 34} \Big)\rangle_t  + 6 \langle
Q_{l_\eta 12}P_{l_\chi 13} Q_{l_\eta 45}P_{l_\chi 45} \rangle_t.
\end{eqnarray}
By assuming a Gaussian behavior, as in the strategy outlined in \cite{sumrule}, we can write the overall  streaming of the correlation
functions in the form of  the following differential system
\begin{eqnarray}
  \dot{A} &=& 2AD - 8BE + 6CF, \nonumber\\
  \dot{B} &=& 2AE + 2BD -
4BE - 6BF - 6EC + 12CF, \nonumber \\
  \dot{C} &=& 2AF + 2CD + 8BE -
16BF - 16CE + 20CF, \nonumber\\
  \dot{D} &=& AG - 4BH + 3CI + D^2
-4E^2 + 3F^2, \nonumber \\
  \dot{E} &=& AH+BG -2BH -3BI -3CH + 6CI
+ 2ED -2E^2 -6EF + 6F^2,\nonumber \\
  \dot{F} &=& AI + CG + 4BH
-8BI -8 CH + 10 CI  + 2DF + 4E^2 -16EF + 10F^2,\nonumber \\
  \dot{G} &=& 2GD - 8HE + 6IF, \nonumber\\
  \dot{H} &=& 2GE + 2HD -
4HE - 6HF -6IE + 12IF, \nonumber\\
  \dot{I} &=& 2GF + 2DI + 8HE -
16 HF - 16IE + 20IF.
\end{eqnarray}
As we are interested in discussing criticality and not the whole glassy phase, it is possible to solve this system starting from the high noise region, once the initial conditions at $t=0$ are known.
As at $t=0$ everything is factorized, the only needed check
is by the correlations inside each party.
Starting with the first party, we have to study $A,B,C$ at $t=0$.
As only the diagonal terms give non-negligible contribution, it is
immediate to work out this first set of starting points as

\bea\nonumber
 &&A(0) =\< Q_{l_\eta 1 2}^2 \>= N  (\< (q^{l_{\eta}}_{12})^2\>-2\E_{l_\eta}\overline{q}^{l_{\eta}}\<q^{l_{\eta}}_{12}\>+\E_{l_\eta}\overline{q}^{l_{\eta}}=\\
 && \qquad = N \big(\E_{l_\eta} \frac{1}{l_{\eta}^2} \<\sum_i^{l_\eta} (\sigma_i^1)^2 (\sigma_i^2)^2 \> + \overline{q}^2 \big)=\left( \frac{1+a}{2}\right)^{-\frac{1}{2}} -N\overline{q}^2 =  \sqrt{\frac{2}{\gamma}}N^{\frac{\theta}{2}}-N\overline{q}^2,\\ 
 &&B(0) = \< Q_{l_\eta 1 2}Q_{l_\eta 1 3} \>= N  (\< q^{l_{\eta}}_{12}q^{l_{\eta}}_{13}\> - \overline{q}^2 )=  \sqrt{\frac{2}{\gamma}}N^{\frac{\theta}{2}} \overline{q}- N\overline{q}^2,\\  \nonumber 
 &&C(0) = \< Q_{l_\eta 1 2}Q_{l_\eta 3 4} \>= N  (\< q^{l_{\eta}}_{12}q^{l_{\eta}}_{34}\> - \overline{q}^2 )= N \big(\E_{l_\eta} \frac{1}{l_{\eta}^2} \< \sum_i^{l_\eta} \sigma_i^1 \sigma_i^2\sigma_i^3\sigma_i^4 \>- N\overline{q}^2=\\
 &&\qquad = \sqrt{\frac{2}{\gamma}}N^{\frac{\theta}{2}} \int d\mu({\eta})\mathrm{tanh}^4\bigg(\frac{ \beta \sqrt{\alpha \overline{q} \g}  N^{\theta/2}}{1-\beta \sg N^{\theta/2}(1-\overline{q})} \bigg)-N\overline{q}^2.
\eea

For the second party we need to evaluate $G,H,I$ at $t=0$. The only difference with the first party is that $z_{\mu}^2 \neq 1$ as for the $\sigma$'s.
\bea  \nonumber
 &&G(0) =\< P_{l_\chi 1 2}^2 \>= N  (\< (p^{l_{\chi}}_{12})^2\>-2\E_{l_\chi}\overline{p}^{l_{\chi}}\<p^{l_{\chi}}_{12}\>+\E_{l_\chi}\overline{p}_{l_{\chi}}^2=\\
 &&\qquad= N \big( \E_{l_\chi} \frac{1}{l_\chi^2} \sum_{\mu}^{l_\chi} \< (z^1_\mu)^2\>_G \< (z^2_\mu)^2\>_G- \overline{p}^2 \big)= \big(\alpha \g \big)^{-\frac{1}{2}} N^{\theta/2} ~\omega^2(z^2)-N\overline{p}^2,\\
 &&H(0) = \< P_{l_\chi 1 2}P_{l_\chi 1 3} \>= N  (\< p^{l_\chi}_{12} p^{l_\chi}_{13}\> - \overline{p}^2 )=  \big(\alpha \g \big)^{-\frac{1}{2}} N^{\theta/2}~ \omega(z)~\omega(z^2)-N\overline{p}^2,\\
 &&I(0) = \< P_{l_\chi 1 2}P_{l_\chi 3 4} \>= N  (\< p^{l_\chi}_{12}p^{l_\chi}_{34}\> - \overline{p}^2 )= \big(\alpha \g \big)^{-\frac{1}{2}} N^{\theta/2}~ \omega^2(z)-N\overline{p}^2.
\eea
Now, $\omega(z^2)$ and $\omega(z)$ are Gaussian integrals and can be explicitly calculated as
\begin{eqnarray}
\omega(z) &=& \frac{\int d\mu(z)  z \exp \big(b z \chi+ \frac{d}{2} z^2\big)}{\int d\mu(z) \exp \big(b z \chi+ \frac{d}{2} z^2\big)}= \frac{b \<\chi\>}{1-d}=0,\\
\omega(z^2) &=& \frac{\int d\mu(z)  z^2 \exp \big(b z \chi+ \frac{d}{2} z^2\big)}{\int d\mu(z) \exp \big(b z \chi+ \frac{d}{2} z^2\big)}= \frac{ 1-d+ b^2 \<\chi^2\>}{(1-d)^2}=\\
&=& \frac{1- \beta \sg N^{\theta/2} (1-\overline{q} +\alpha \overline{p})}{\big(1-\beta \sg N^{\theta/2} (1-\overline{q})\big)^2}. \nonumber
\end{eqnarray}
Finally, we have obviously $D(0)=E(0)=F(0)=0$, because at $t=0$ the two parties are independent. As we are interested in finding where ergodicity becomes broken (the critical line),  we start propagating  $t$ (from $0$ to $1$) from the annealed region (high noise limit), where $\bar{q} \equiv 0$ and $\bar{p}\equiv 0$.
It is immediate to check that, for the only terms that we need to
consider, $A,D,G$ (the other being strictly zero on the whole
$t\in [0,1]$), the starting points are:
\bea
&&A(0)=\sqrt{\frac{2}{\gamma}}N^{\frac{\theta}{2}}=\frac{1}{r},\\
&&D(0)=0,\\
&&G(0)=\frac{N^{\frac{\theta}{2}}}{\sg(1-\beta\sg N^{\frac{\theta}{2}})^2}=\frac{1}{r s^2}.
\eea
Where we have defined $r=\sg N^{-\frac{\theta}{2}}$, $s=1-\beta \sg N^{\frac{\theta}{2}}$.

The evolution is ruled by
\bea
\dot{A}=2 A D\\
\dot{D}= A G+ D^2\\
\dot {G}=2G D.
\eea
Noticing that $\frac{\dot{A}}{\dot{G}}=0$ by substitution, and that $\frac{A(0)}{G(0)}=s^2$ we obtain immediately :
\beq\label{AG}
A(t)=G(t)s^2= G(t) \bigg(1-\beta \sg N^{\frac{\theta}{2}}\bigg)^2 .\\
\eeq
The system then reduces to two differential equations;  calling $Y=D+G s$, we have $\dot{Y}=\dot{D}+\dot{G}s= G^2s^2+D^2+2GDs=Y^2$ with solution $Y(t)=\frac{Y(0)}{1-tY(0)}$, and
$Y(0)=D(0)+G(0)s=\frac{1}{r s}$ by which we get
\beq
  Y(t= \sqrt{\alpha} \beta \frac{\gamma}{2})=\frac{1}{r s} ~ \frac{1}{1- \sqrt{\alpha} \beta \frac{\gamma}{2} (r s)^{-1}}=\frac{1}{\sg N^{-\theta/2}  \bigg(1-  \beta \sg N^{\theta/2} (1+\sqrt{\alpha})\bigg)},
\eeq
i.e. there is a regular behavior up to
\beq
\beta_c=\frac{1}{\sg N^{\frac{\theta}{2}} (1+\sqrt{\alpha})}
\:,
\eeq
which confirms the result obtained in Equation (\ref{critica1}).
Now, we can consider separately the evolution equation for $G$ and $D$:
\beq
\dot{G}= 2G(t) \bigg( Y(t) - s G(t)\bigg)= \frac{2}{r s -t} G(t)- 2s G(t),
\eeq
where we used $Y(t)=(r s -t)^{-1}$. Dividing both sides by $G^2$ and calling $Z=G^{-1}$ we get an ordinary first order differential equation for $Z(t)$:
\beq
- \dot{Z}(t)=2Y(t) Z(t)-2 s= \frac{2}{r s -t} Z(t)-2s.
\eeq
that have the following solution for the initial condition $Z(0)=r s^2$:
\beq
Z(t)=2s (r s-t)-\frac{1}{r} (r s-t)^2.
\eeq
From $Z(t)$ we obtain $G(t)$, that is,
\beq
G(t)=\frac{1}{(rs-t)(s+ \frac{t}{r})}.
\eeq
Using Equation (\ref{AG}) and remembering that $D(t)=Y(t)-G(t)s$, we obtain the other overlap fluctuations
\bea
  &&\<Q_{l_\eta 12}^2\>=\frac{\big(1-\beta \sg N^{\theta/2}\big)^2}{ \sg N^{-\theta/2}  \bigg(1-  \beta \sg N^{\theta/2} (1+\sqrt{\alpha})\bigg)\bigg(1- \beta \sg N^{\theta/2}(1-\sqrt{\alpha})\bigg)},\\
  &&\<Q_{l_\eta 12} P_{l_\chi 12}\>=\frac{\sqrt{\alpha} \beta}{ \bigg(1-  \beta \sg N^{\theta/2} (1+\sqrt{\alpha}) \bigg) \bigg(1- \beta \sg N^{\theta/2}(1-\sqrt{\alpha}) \bigg)},\\
  &&\<P_{l_\chi 12}^2\>=\frac{1}{ \sg N^{-\theta/2}  \bigg(1-  \beta \sg N^{\theta/2} (1+\sqrt{\alpha})\bigg)\bigg(1- \beta \sg N^{\theta/2}(1-\sqrt{\alpha}) \bigg)}.
\eea
A simple visual inspection of the formula above allows to confirm that the poles are located at
$$
\beta~\sg ~N^{\theta/2}~(1+\sqrt{\alpha})=1,
$$
confirming the heuristic result previously obtained.
We can easily see furthermore that in the fully connected limit ($\gamma=2$ and $\theta=0$) we recover the result of \cite{amit}.

\section{Conclusions and outlooks}

In this paper we introduced and solved, at the replica symmetric level, two disordered mean-field systems: the former provides a  generalization of the analogue neural network by introducing dilution into its patterns encoding the memories, the latter is a bipartite and diluted spin-glass made up of a Gaussian party and an Ising party, respectively.
From an applicative viewpoint (not discussed here, see e.g. \cite{galluzzi}), the interest in these models raises in different contexts, but their peculiarity resides in the existence of sparse entries (instead of classical dilution on the neural network links as performed for instance  earlier by Sompolinsky \cite{sompolinsky} through random graphs or recently by Coolen and coworkers \cite{ton1} through small-worlds or scale-free architectures) which allows, when possible, parallel retrieval as for instance discussed in \cite{galluzzi}. Interestingly, as we show, the Hamiltonians describing these systems are thermodynamically equivalent.

In our investigations we first considered the diluted analogue neural network and focused on the topological properties of the emergent weighted graph. We found an exact expression for the coupling distribution, showing that in the thermodynamic limit it converges to a central Gaussian distribution with variance scaling linearly with the system size $N$. We also calculated the average link probability which, as expected, depends crucially on the degree of dilution introduced. More precisely, by properly tuning it, the emergent structure displays an average coordination number $\bar{z}$ which can range from $\bar{z} = N$ (fully-connected regime) to $\bar{z} = \mathcal{O}(N)$ (constant link probability), to finite with $\bar{z}>1$ (overpercolated network) or $\bar{z}<1$ (underpercolated network).

Then, we moved to the thermodynamical analysis, where, through an interpolation scheme recently developed for fully connected Hebbian kernels \cite{BGG1}, we obtained explicitly the replica symmetric free-energy coupled with its self-consistency equations. The overlaps, order parameters of the theory, turn out to be classical weighted sums of sub-overlaps defined on all possible sub-graphs (as for instance discussed in \cite{agliari2,ton2}). Both a small overlap expansion of these self-consistencies, as well as a whole fluctuation theory developed for their rescaled correlations, confirm a critical behavior on a surface (in the $\alpha,\beta,\gamma,\theta$ iperplane) that reduces to the well-known of Amit-Gutfreud-Sompolinsky when the dilution is sent to zero \cite{amit}. On the other hand, the net effect of  entry dilution in bitstrings (which weakens the coupling strength) is to rescale accordingly the critical noise level at which ergodicity breaks down, as expected.
\newline
Without imposing retrieval through Lagrange multipliers (as for analogue patterns it is not a spontaneous phenomenon, see  \cite{BGG1}) the system displays only two phases, an ergodic one (where all overlaps are zero) and a spin-glass one (where overlaps are non-zero), split by the second order critical surfaces (over which overlaps start being non-zero) which defines criticality.

Outlooks should follow in two separate directions: from a pure speculative and modeling viewpoint  attention should be paid to the mechanism of replica-symmetry breaking, which is known to happen on these models, while for an applicative perspective they could be integrated in the theoretical framework where collocate the plethora of experimental results stemmed from complex system analysis, e.g. from neural and immunological contexts.
We plan to report soon on both the topics.

\section*{Acknoledgments}
This work is supported by the MiUR through the FIRB project RBFR08EKEV and by Sapienza Universit\`a di Roma.\\
The authors acknowledge Anton Bovier for a fruitful comment.

\bibliographystyle{unsrt}
\bibliography{bibliografia}

\end{document}